\documentclass[aps,pra,twocolumn,superscriptaddress,notitlepage,nofootinbin]{revtex4-1}
\usepackage{mathtools}
\usepackage{amsfonts}
\usepackage{latexsym}
\usepackage{relsize}
\usepackage{euscript}
\usepackage{amssymb}
\usepackage{graphicx}
\usepackage{amsmath}
\usepackage{amsbsy}
\usepackage{amsthm}
\usepackage{bbm}
\usepackage{bm}
\usepackage{epsfig}
\usepackage{epstopdf}
\usepackage{dsfont}

\usepackage[colorlinks]{hyperref}
\usepackage[figure,table]{hypcap}
\usepackage{enumerate}
\usepackage{geometry}
\hypersetup{
	bookmarksnumbered,
	pdfstartview={FitH},
	citecolor={darkgreen},
	linkcolor={darkred},
       linktoc={page},
	urlcolor={darkblue},
	pdfpagemode={UseOutlines}}

\usepackage{color}

\definecolor{darkgreen}{RGB}{40,130,40}
\definecolor{darkblue}{RGB}{0,0,190}
\definecolor{darkred}{RGB}{238,0,0}
\usepackage{soul}
\usepackage{float}
\usepackage{algorithm}

\floatname{algorithm}{Protocol}

\def\EQ#1{\begin{eqnarray}#1\end{eqnarray}}
\newcommand{\djj}{d\kern-0.4em\char"16\kern-0.1em}

\newtheorem{prop}{Proposition}\def\PRO{\begin{prop}}\def\ORP{\end{prop}}
\newtheorem{coro}{Corollary}\def\COR{\begin{coro}}\def\ROC{\end{coro}}
\newtheorem{theo}{Theorem}\def\TH{\begin{theo}}\def\HT{\end{theo}}
\def\TH{\begin{theo}}\def\HT{\end{theo}}
\newtheorem{defi}[prop]{Definition}\def\DE{\begin{defi}}\def\ED{\end{defi}}
\newtheorem{lemme}[prop]{Lemma}\def\LE{\begin{lemme}}\def\EL{\end{lemme}}

\def\one{\mathbb{I}}

\def\ket#1{\left| #1 \right\rangle}
\def\bra#1{\left\langle #1 \right|}
\def\dm#1{\left|#1 \right\rangle \left\langle #1 \right|}

\begin{document}
\title{Quantum speedup for active learning agents}

\author{Giuseppe Davide Paparo$^1$}
\author{Vedran Dunjko$^{2,3,4}$}
\noaffiliation
\author{Adi Makmal$^{2,3}$}
\noaffiliation
\author{Miguel Angel Martin-Delgado$^1$}
\noaffiliation
\author{Hans J. Briegel$^{2,3}$\\ \vspace{0.3cm}}
\noaffiliation
\affiliation{Departamento de Fisica Teorica I, Universidad Complutense, 28040 Madrid, Spain\\
$^2$Institut f\"{u}r Theoretische Physik, Universit{\"{a}}t Innsbruck, Technikerstra{\ss}e 25, A-6020 Innsbruck, Austria\\
$^3$Institut f\"{u}r Quantenoptik und Quanteninformation der {\"{O}}sterreichischen Akademie der Wissenschaften, A-6020 Innsbruck, Austria\\
$^4$Division of Molecular Biology, Ru\djj er Bo\v{s}kovi\'{c} Institute, Bijeni\v{c}ka cesta 54, 10002 Zagreb, Croatia}


\begin{abstract}

Can quantum mechanics help us build intelligent learning agents? A defining signature of intelligent behavior is the capacity to learn from experience. However, a major bottleneck for agents to learn in real-life situations is the size and complexity of the corresponding task environment. Even in a moderately realistic environment, it may simply take too long to rationally respond to a given situation. If the environment is impatient, allowing only a certain time for a response, an agent may then be unable to cope with the situation and to learn at all. Here we show that quantum physics can help and provide a quadratic speed-up for active learning as a genuine problem of artificial intelligence.  This result will be particularly relevant for applications involving complex task environments.
\end{abstract}

\titlepage
\maketitle
\section{Introduction}

The levels of modern day technology have, in many aspects, surpassed the predictions made in the mid-$20^{th}$ century, as is easily witnessed, for example, by the sheer computing power of the average `smart' mobile phone.
Arguably, the most striking exception to this, apart from, perhaps, human space exploration, lies in the development of genuine artificial intelligence (AI), the challenge of which has initially been greatly underestimated. The unceasing setbacks in the general AI problem caused research to shift emphasis to the production of useful technology, a direction now called applied AI. That is, emphasis was placed to specific algorithmic AI tasks -- modules, such as data clustering, pattern matching, binary classification, and similar -- and reduced from the holistic task of designing an autonomous and intelligent agent.

The discovery that the laws of quantum physics can be employed for dramatically enhanced ways of information processing \cite{1985_Deutsch,1992_Deutsch,1996_Grover,1994_Shor,2000_NC,2000_Bennet} has already had a positive influence on specific algorithmic tasks of applied AI  \cite{2002_Sasaki,2008_Neven,2013_Lloyd,2009_Brukner,2013_Lidar,2013_Aimeur}.
However, to our knowledge, it has so far not been demonstrated that quantum physics can help in the complemental task of designing autonomous and learning agents. 
The latter task is studied in the fields of embodied cognitive sciences and robotics \cite{1986_Braitenberg, 1999_Brooks, 1999_Pfeifer, 2006_Pfeifer, 2008_Floreano, 2008_Barsalou}, which promote a behavior-based approach to intelligence and put a strong emphasis on the physical aspects of an agent.
The approach to AI we adopt in this work is along the lines of the latter perspective. We are guided by a few basic principles, inspired by biological agents, which include autonomy (implying that the agent must learn in, and adapt to, unknown dynamic environments), embodiedness (implying that the agent is situated in, and actively interacts with, a physical environment), and homogeneity (meaning that all possible separate Òcognitive unitsÓ arise as possible configurations of one, or a few, homogeneous underlying systems that are, in principle, capable of growth). An example of a model that one could consider homogenous, aside from the projective simulation model  \cite{2012_Briegel} we will consider here, would for example be artificial neural networks. One may then envision that true AI will emerge by growth and the learning of an agent, rather than through deliberate design. 
In this paper, we show that in such an embodied framework of AI provable advancements of a broad class of learning agents can be achieved when we take the full laws of quantum mechanics into account.

\section{Learning agents and quantum physics}
\begin{figure}
\label{fig:AgentBasic}
\includegraphics[width=0.48\textwidth]{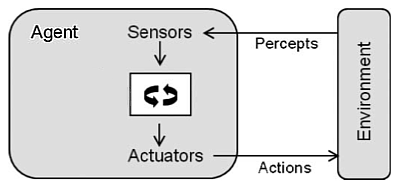}
\caption{
An agent is always situated in an environment. It is equipped with \emph{sensors}, through which it receives \emph{percepts} from the environment, and with \emph{actuators}, through which it can act on the environment. Based on perceptual input, the agent will after some internal processing engage its \emph{actuators} and output an \emph{action}. Adapted and modified from~\cite{2003_Russel}.}
\end{figure}

How could quantum physics help design better agents?
An embodied agent is always situated in an environment from which it receives a sensory input, that is, a \emph{percept} (from some set of percepts $\mathcal{S} = \{s_1, s_2 \ldots \}$)  and, based on the percept, it produces an action from the possible set of actions $\mathcal{A} = \{a_1, a_2, \ldots \}$, see Fig.~\ref{fig:AgentBasic}.
The capacity to learn implies that the agent is at every instant of time in some \emph{internal state}, that can change based on previous sequences of percept-action events. That is, it has \emph{memory} which reflects the agents history. 
The typical model for such autonomous learning agents we consider here is the reinforcement learning model \cite{2003_Russel, SuttonBarto98}, where to each percept-action event a reward in $\Lambda =\{0,1 \}$  (for simplicity, we consider binary rewards,  but this reward system can be easily generalized in our model) is assigned when the action was correct.

Each percept-action-reward sequence constitutes an external time-step (or cycle) of the activity of an agent.
The learning process of the agent is characterized by an update rule of the internal state (based on the previous percept-action-reward sequences), and the (local) policy of the agent  \cite{SuttonBarto98} is defined by what action is output given the current internal state and the received percept. Unlike in typical reinforcement learning settings, in embodied active agents the required time to evaluate the policy (decide on an action) must be taken into account, and we refer to it as \emph{internal time}.

The agent's learning process is reminiscent to computational \emph{oracle query models}, in which an unknown oracle (environment), is queried (via an action) by the agent, in an iterative quest for the best responses.  It is tantalizing to consider employing the powerful quantum searching machinery \cite{1996_Grover, 2004_Szegedy_IEEE, 2011_Magniez_SIAM}, which has been proven to outperform classical algorithms in computational settings, in an attempt to improve the agent. 

However, contrary to computer algorithms, an embodied agent, such as a robot, operates 
in a physical environment
which is, for most existing applications, classical \footnote{{Examples of such applications include the problems of navigation, or human-robot interaction. We note that the focus on such classical environments is by no means a restriction of our scheme. 
For a quantum environment, the actions which the agent can perform could e.g. be quantum measurements (as components of his actuators), and percepts the measurement outcomes -- such scenarios are certainly not without interest. 
The results we present in this work apply equally to such environments.
However, we will not explore the applications of learning agents in quantum environments in this paper.}}. 
This prohibits querying in \emph{quantum superposition}, a central ingredient to all quantum search algorithms. Thus, such na\"{i}ve approaches to quantizing learning agents are doomed to fail \footnote{Even if we were to allow superpositions of actions, the amount of control the agent must have over the degrees of freedom of the environment, in order to apply quantum query algorithms, may be prohibitive. This constitutes one of the fundamental operative distinctions between \emph{quantum algorithms,} where full control is assumed, and \emph{quantum agents,} where such control is limited.}.

Nonetheless, while the physical nature of the agent and the environment prohibits speed-up through quantum queries of the environment, the physical processes \emph{within} the agent, which lead to the performed actions, can be significantly improved by employing full quantum mechanics \footnote{In embodied agents, these physical processes may \emph{e.g.} realize some internal representation of the environment, which the agent itself has to develop as it interacts with the environment. For example, in the context of artificial neural networks such internal models are known as self-organizing maps and, more specifically, sensorimotor maps \cite{1995_Kohonen,2006_Toussaint}.}.
In particular, the required internal time can be polynomially reduced in the model we present next. 
In general learning settings, this speedup alone will constitute an overall qualitative improvement of performance, 
for instance when the environment changes on time-scales not overwhelmingly larger than the agent's internal `thinking' time.
\section{Quantum agents based on projective simulation}

\subsection{The PS agent model}
The AI model of the agents we consider in the following is the so-called Projective Simulation (PS) model \cite{2012_Briegel}, whose conceptual framework is in line with the desired guiding principles we highlighted earlier.
The PS model is based on a specific memory system, which is called episodic and compositional memory (ECM). This memory provides the platform for \emph{simulating future action} before real action is taken. The ECM can be described as a stochastic network of so-called \emph{clips}, which constitute the elementary excitations of episodic memory and can be implemented as excitations of suitable physical systems.
The percepts  ($\mathcal{S}$) and actions ($\mathcal{A}$), along with sequences thereof, are represented within an agent as the aforementioned clips, and the set of these comprises the \emph{clip space} $C = \{ c = (c^{(1)}, c^{(2)}, \ldots) \vert c^{(k)} \in  \mathcal{S} \cup \mathcal{A}\}$. 
In this work we consider clips which are unit length sequences, representing a \emph{memorized} percept or an action, which we denote using the same symbols, so $C =  \mathcal{S} \cup \mathcal{A}$ \footnote{In the PS framework one formally distinguishes between real percepts and actions $s,a$, and their internal representations denoted $\mu(s),\mu(a)$, which comprise clips. For our purposes, by abuse of notation, we will omit the mapping $\mu$. For more details see the Appendix, section \ref{abst:ThePSModel}. }, but this can be easily generalized. The internal states of the agent, i.e. the total memory,
comprise weighted graphs $G_s$ over subsets of the clip space, which are assigned to each percept $s$. This graph dictates the hopping probabilities from one clip to another, and the hopping process realizes a Markov chain (MC).  Thus, the elementary internal processes of the agent, which implement the transitions from clip to clip, are \emph{discrete-time stochastic diffusion processes}. These diffusion processes can be realized in a variety of physical systems, as we discuss later. 

The deliberation process of the agent is based only on the diffusion processes over the clip space of a certain (and, in general, variable) size, making this model homogeneous in the sense we explained earlier.
The PS agents also perceive rewards $\Lambda=\{0,1\}$ and, based on the percieved percept $s$, realized action $a$, and the resulting reward, the weights of the graph $G_s$ (that is, transition probabilities) are updated via simple rules~\footnote{In the PS, the hopping probabilities themselves are encoded in the so-called $h-$matrix, which is an un-normalized representation of the transition matrix \cite{2012_Briegel}.}. 

While the PS framework allows for many additional structures (see the Appendix, section \ref{abst:ThePSModel} for further details on the PS model),  we will, for simplicity, only consider  percept-specific \emph{flags} -- corresponding to rudimentary \emph{emoticons} in \cite{2012_Briegel} -- which are subsets of actions assigned to each percept, formally $\mathcal{F} = \{f(s) \vert f(s) \subseteq \mathcal{A}, s\in \mathcal{S} \}$. Such flags may be used to represent the agent's short-term memory, in which case they significantly improve the performance of the model \cite{2012_Briegel}.
For example, one can consider a possible mechanism in which for each percept, all actions are initially flagged. If the agent outputs some action $a$, given a percept $s$, and this action is not rewarded, $a$ is removed from $f(s)$. Once the set $f(s)$ has been depleted (indicating, for instance, that the environment changed its policy), it is re-set to contain all actions.
The meaning of flags may be more general, and in this work we only assume the sets of flags are always non-empty.

In the process of deliberation, the relevant Markov chain is diffused a particular number of times, depending on the particular PS model, until an action is output via so-called output couplers \footnote{Each agent is equipped with input and output couplers, which translate, through sensors and actuators (see Fig.~\ref{fig:AgentBasic}), real percepts to the internal representations of percepts, and internal representations of actions to real actions. }. 
The choice of the action - and thereby the policy of the agent - is dictated by the probability distribution over the clip space, which is realized by the diffusion process. The latter depends on the agent's experience manifest in the specified MC. Intuitively, this distribution represents \emph{the agent's state of belief} on what is the right action in the given situation.

A particular model of PS we introduce here, so-called  \emph{reflecting} PS (r-PS) agents, draw their name from the reflection process \cite{2012_Briegel} in which the diffusion processes are repeated many times. Such agents approximate the complete mixing of their MCs, simulating infinite deliberation times. Once the mixing is (approximately) complete, the reflecting agent samples from the realized stationary distribution over the clip space (and, if needed, iterates the mixing process) until a flagged action clip has been sampled.
The internal states of reflecting PS agents are thus \emph{irreducible, aperiodic and reversible MCs} over subsets of the clip space (which contain all action clips).
Reflecting PS agents can be seen as a generalization of so-called standard PS agents, and a comparison between a well-studied class of PS agents, and r-PS agents is provided in the Appendix, section \ref{subsect:comparison}.

In the limit of complete mixing, given a percept $s$ and the current internal state (the MC $P_s$), the r-PS agents output an action $a$ distributed according to $\tilde{\pi}_s$ given as follows: let $\pi_s$ be the stationary distribution of $P_s,$ and let $f(s)$ be the subset of flagged actions, then
\EQ{
\tilde{\pi}_s(i) = \left\lbrace \begin{tabular}{cl}\vspace{0.1cm}
$\dfrac{\pi_s(i)}{\sum_{j \in f(s)} \pi_s(j)},$& $\textup{if} \ i\in f(s)$\\
0,& \textup{otherwise},
\end{tabular}  \right. \label{EQ3}
}
that is the re-normalized stationary distribution $\pi_s$ modified to have support only over flagged actions. We will often refer to $\tilde{\pi}_s$ as \emph{the tailed distribution}.

In general, complete mixing is not possible or needed.
To realize the (approximate) tailed distribution, as given in Eq. \ref{EQ3}, the classical agent will have to, iteratively, prepare the approximate stationary distribution of $P_s$ (by applying $P_s$ to the initial distribution $t_{mix}^c$ times), and sample from it, until a flagged action is hit. 
It is well known that, in order to mix the MC, $t_{mix}^c$ should be chosen in $\tilde{O}( 1/\delta_{s})$
\footnote{In this paper we do not consider logarithmically contributing terms in the complexity analysis, thus we use the $\tilde{O}$-level analysis of the limiting behavior (instead of the standard 'big O' $O$).} (where $\delta_s$ is the spectral gap of MC $P_s$ defined as $\delta_s = 1-|\lambda_2|$ and $\lambda_2$ is the second largest eigenvalue of $P_s$ in absolute value),
and the expected number of iterations of mixing and checking which have to be performed is $t_{check}^c \in \tilde{O}(1/\epsilon_{s})$ (where $\epsilon_{s} = \sum_{i\in f(s)} \pi(i)$ is the probability of sampling a flagged action from the stationary distribution $\pi_s$). Here, we will use the label $P_s$ of the transition matrix as a synonym for the MC itself.
Note that the internal time of the classical agent, i.e. the number of primitive processes (diffusion steps), is therefore governed by the quantities $1/\delta_{s}$ and $1/\epsilon_{s}$.

\subsection{Quantum speed-up of reflecting PS agents}
The procedure the r-PS agent performs in each deliberation step resembles a type of a random walk-based search algorithm which can be employed to find targeted items in directed weighted graphs  \cite{2011_Magniez_SIAM}. In that context, the theory of quantum walks provides us with analogs of discrete-time diffusion processes, using which the search time can be quadratically reduced \cite{2011_Magniez_SIAM,2005_Magniez, 2004_Ambainis,2004_Szegedy_IEEE}. 
To design the quantum agent, inspired by these approaches to searching, here we introduce a quantum walk procedure which can be seen as a randomized Grover-like search algorithm (the quantum counterparts of classical search algorithms), most closely matching the main protocol in \cite{2011_Magniez_SIAM}. 

However, there are essential differences between searching problems and the problems of designing intelligent AI agents, which we expose and resolve in this work.
First, we note that, for the task of simple searching, the procedure the r-PS agent follows is known not to be optimal in general \cite{2011_Magniez_SIAM, 2012_Magniez_Algorithmica}. In contrast, for the task of the r-PS agent, which is to output a flagged action \emph{according to a good approximation of the tailed distribution} in Eq. \ref{EQ3}, this algorithm is, in general, optimal \footnote{The optimality claim holds, provided that no additionaly mechanisms except for diffusion and checking are available.}. This can be seen by the known lower bounds for mixing times $t^{c}_{mix}$ of reversible MCs (see the Appendix, section \ref{subsubsect:(Q)Wbasics} for details).  

Furthermore, while as a direct consequence of the results in \cite{2011_Magniez_SIAM}, the quantum r-PS produces a flagged action in times quadratically faster than is achieved using the procedure employed by the classical agent, prior works provide no guarantees that the output actions will be distributed according to the desired tailed  distribution. Recall, in the context of AI, all agents produce some action, and it is precisely the output distribution which differentiates one agent from another in terms of behavior (and, thus success).
 In this work we prove that both the output distributions of the reflecting classical and quantum agents approximate the distribution of Eq. \ref{EQ3}, and thus are approximately equal. 
    That is, they belong to the same \emph{behavioral class} (for a formal definition of behavioral classes we introduce see the Appendix, section \ref{subsect:Formaldefinitionsofreflectingagents}). Consequently, the quantum reflecting agent construction we give realizes, in full sense, quantum-enhanced analogs of classical reflecting agents.

While quantum approaches to sampling problems have not until now been exploited in AI, we observe that the methodology we use is, in spirit, related to the problem of sampling from particular distributions. Such sampling tasks have been extensively studied, often in the context of Markov chain Monte Carlo methods, where quantum speed-up can also be obtained \cite{2011_Temme_Nature, 2012_Yung, 2008_Somma_P,  2008_Wocjan}.
There the quantum walks were mostly used for the important purpose of sampling from Bolzmann-Gibbs distributions of (classical and quantum) Hamiltonians.

In order to define the quantum reflecting agent, we first review the standard constructions and results from the theory of classical and quantum random walks  \cite{2004_Ambainis,2004_Szegedy_IEEE, 2011_Magniez_SIAM}, and refer the reader to the Appendix, section \ref{subsubsect:(Q)Wbasics}, for more details.

The quantum r-PS agent we propose uses the standard quantum discrete time diffusion operators $U_{P_s}$ and $V_{P_s}$ which act on two quantum registers, sufficiently large to store the labels of the nodes of the MC $P_s$.
The diffusion operators are defined as
$U_{P_s}\ket{i}\ket{0} = \ket{i} \ket{p_i}$ and $V_{P_s} \ket{0}\ket{j} = \ket{p^*_j} \ket{j}$
where $\ket{p_i} =  \sum_j \sqrt{[P_{s}]_{ji}}\ket{j}$, $\ket{p^*_j} =\sum_i \sqrt{[P^*_s]_{ij}}\ket{i}$. Here, $P^*_s$ is the time reversed MC  defined by $\pi_i P_{ji} = \pi_j P^*_{ij}$, where $\pi = (\pi_i)_i$ is  the stationary distribution of $P_s$ \footnote{In the case of reversible MCs, which will be the main focus of this paper, $P=P^*$, so $V$ can be constructed from $U$ by conjugating it with the swap operator of the two registers. Here, we present the construction for the general case of irreducible, aperiodic Markov chains.}.
Using four applications of the diffusion operators above, it has been shown that one can construct the standard \emph{quantum walk operator} $W(P_s)$ (sometimes referred to as the Szegedy walk operator), which is a composition of two reflections in the mentioned two-register state space. In particular, let $\Pi_1$ be a projection operator on the space $\mathrm{Span}\{\ket{i} \ket{p_i}\}_i$ and $\Pi_2$ be the projector on $ \mathrm{Span} \{\ket{p^*_j} \ket{j} \}_j$.
Then $W(P_s) = (2\Pi_2- \one) (2\Pi_1- \one)$.

{Using the quantum walk operator and the well-known phase detection algorithm \cite{2000_NC,2011_Magniez_SIAM} the agent can realize the $R(P_s)(q,k)$ subroutine which approximates the reflection operator $2 \dm{\pi_s} - \mathbbmss{1},$ where $\ket{\pi_s} = \sum_{i} \sqrt{\pi_s(i)} \ket{i}$ is the coherent encoding of the stationary distribution $\pi_s$. The parameters $q$ and $k$ control the fidelity (and the time requirement) of this process, i.e. how well the reflection is approximated as a function of the number of applications of the quantum walk operator. 

{More precisely, in the implementation of the $R(P_s)(q,k)$ operator, the quantum r-PS agent utilizes an ancillary register of $q\times k$ qubits. To ensure the correct behavior, $q$ is chosen as $q\in \tilde{O}(1/\sqrt{\delta_{s}})$, which depends on the square root of the spectral gap of the MC $P_s$.}
{Under this condition, it has been shown that the distance between the ideal reflection operator and the realized approximate operator is upper bounded by $2^{1-k}$, under a suitable metric (see the Appendix,  section \ref{subsubsect:(Q)Wbasics}, Theorem 3  for details).}
That is, the fidelity of this reflection operator approaches unity exponentially quickly in the parameter $k$ \cite{2011_Magniez_SIAM}.

To produce a flagged action according to the desired distribution, the quantum r-PS agent will first initialize its quantum register to the state $\ket{\pi_{init}} = U_{P_s} \ket{\pi_s} \ket{0}$ which requires just one application of the diffusion operator $U_{P_s},$ provided the state $\ket{\pi_s} = \sum_{i} \sqrt{\pi_s(i)} \ket{i}$ is available.
Here, like in the standard frameworks of algorithms based on quantum walks with non-symmetric Markov chains, we assume that the state $\ket{\pi_s}$ is available, and in the Appendix, section \ref{subsect:comparison}, we provide an example of reflecting classical and quantum agents where this is easily achieved \footnote{Here, we note that concrete applications of quantum walks specified by non-symmetric Markov chains have, to our knowledge and aside from this work, only been studied in \cite{2012_paparo_google,2013_paparo_complex}, by two of the authors and other collaborators, in a significantly different context.}. 

Following this, the agent performs a randomized Grover-like sequence of reflections, reflecting over flagged actions (denoted $ref({f(s)})$), interlaced with reflections using the approximate reflection operator described previously. 
After the reflections have been completed the required number of times, the resulting state is measured, and the found flagged action is output. In the case a non-action clip is hit, the entire procedure is repeated \footnote{For completeness we note as a technicality, following \cite{2011_Magniez_SIAM}, that if $\epsilon_s$ is always bounded below by a known constant (say $3/4$ as in \cite{1998_Boyer}), the quantum agent can immediately measure the initial state and efficiently produce the desired output by iterating this process a constant number of times. However, in the scenarios we envision, $\epsilon_s$ is very small. }.

Since Grover-like search engines guarantee that the overlap between the final state, and a state with support just over the actions is constant, this implies that the probability of not hitting a flagged action decreases exponentially quickly in the total number of iterations, and does not contribute significantly to our analysis.  In the Appendix, section \ref{subsect:Formaldefinitionsofreflectingagents}, we provide a detailed analysis, and propose a method for the efficient re-preparation of the required initial state (by recycling of the residual state), in the event the deliberation procedure should be repeated. 
The deliberation process is detailed in Fig. \ref{qagent}.
The total number of reflections required is $t_{check}^q \in_{R} [0,\tilde{O}(1/\sqrt{\epsilon_{s}})]$ (this choice is uniform at random as for the randomized Grover algorithm \cite{1998_Boyer}), and the approximate reflection operator is applied with parameters $R(P_s)(t_{mix}^q,\tilde{O}(1)),$ where $t_{mix}^q \in \tilde{O}( 1/\sqrt{\delta_{s}})$ \footnote{We note that some of the only logarithmically contributing terms, which appear in a more detailed analysis omitted here, can be further avoided using more complicated constructions as has, for instance, been done in \cite{2011_Magniez_SIAM}.}.

This implies that the total number of calls to the diffusion operators $U_{P_s}$ and $V_{P_s}$ is in $\tilde{O}(1/(\sqrt{\epsilon_{s} \delta_s})$ (and the total number of reflections over flagged actions -- equivalents of checks in the classical agent -- is in $\tilde{O}(1/\sqrt{\epsilon_s})$), which is a quadratic improvement over the classical agent.
As we have mentioned previously, the remaining key ingredient to our result is the fact that the proposed  quantum agent produces actions (approximately) according to the tailed distribution in Eq. \ref{EQ3}, and that the approximations for both agents can be made arbitrarily good within at most logarithmic overhead. The proof of this claim we leave for the Appendix section \ref{subsect:twomain}. We note that in this paper we have presented constructions for reversible MCs for simplicity, but the constructions can be extended to general irreducible chains using approaches analogous to those in  \cite{2011_Magniez_SIAM}.

We have thus presented a method for generic quantization of reflecting PS agents, which maintains the behavior of the agents, and provably yields a quadratic speed-up in internal times vital in real environment settings.

\begin{figure}[h!]
\includegraphics[trim =3cm 11cm 12cm 5cm ,width=0.48\textwidth,clip=true]{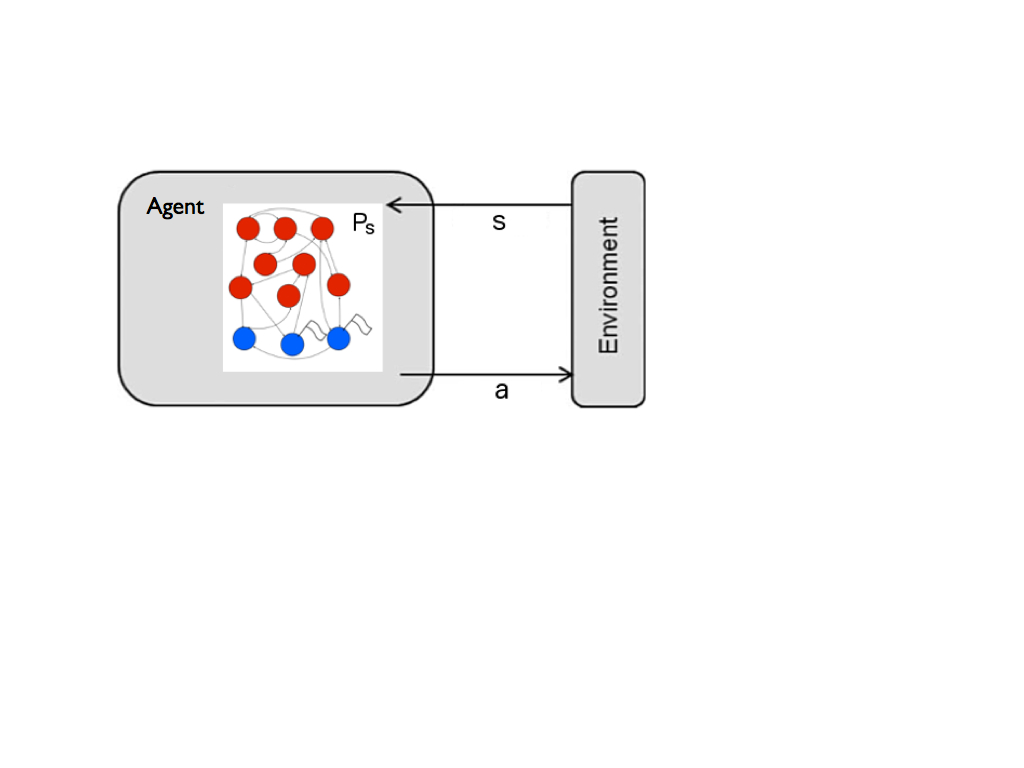}
\label{qagent}
\begin{footnotesize}
\hrule
\begin{enumerate}
\item \textbf{Initialize:} Prepare $$\ket{\pi_{init}} = \sum_{i\in X} \sqrt{\pi_s(i)} \ket{i} \ket{p_i},$$ with $\ket{p_i} = \sum_{j} \sqrt{p_{ji}} \ket{j}, $ and $p_{ji}$ is the transition probability from $i$ to $j$ as dictated by the MC $P_s$.
\item For $t_{check}^q$ time-steps do\\
\begin{enumerate}
\item \textbf{Check:} Apply the $ref({f(s)})$ operator which flips the phase of all components of the current state of the first register which are not in $f(s)$.
\item \textbf{Diffuse:} Apply the approximate reflection operator $R(P_s)(t_{mix}^q,O(1))$, as described in main text.
\end{enumerate}
\item Measure the first register, and if it is a flagged action, output it, else reiterate the procedure.
\end{enumerate}
\hrule
\end{footnotesize}
\caption{Both the classical and the quantum r-PS are characterized by their internal Markov chains over the clip space.
The procedures the quantum agent performs in order to choose the desired actions are listed. The values $t_{mix}^q$ and $t_{check}^q$ are chosen in $\tilde{O}(1/\sqrt{\delta_s})$ and uniformly at random in $[0, \tilde{O}(1/\sqrt{\epsilon_s})]$, respectively, see text for details.}
\end{figure}

\section{Discussion}

We have presented a class of quantum learning agents that use quantum memory for their internal processing of previous experience. These agents are situated in a classical task environment that rewards a certain behavior but is otherwise unknown to the agent, which corresponds to the situation of conventional learning agents.  
 
 The agent's internal `program' is realized by physical processes that correspond to quantum walks. These quantum walks are derived from classical random walks over directed weighted graphs, which represent the structure of its episodic memory. We have shown how, using quantum coherence and known results from the study of quantum walks, the agent can explore its episodic memory in superposition in a way which guarantees a provable quadratic speedup in its active learning time over its classical analogue.

 Regarding potential realizations for such quantum learning agents, modern quantum physics laboratories are exploring varieties of systems which can serve as suitable candidates. Quantum random walks and related processes can naturally be implemented in linear optics setups by, for instance, arrays of polarizing beam splitters 
\cite{2012_Aspuru} and highly versatile setups can also be realized using internal states of trapped ions \cite{2012_Roos}. Such advancements, all of which belong to the field of quantum simulation \cite{2013_Schaetz}, could be used as ingredients towards the implementation of quantum reflecting agents, without the need to develop a full-blown universal quantum computer.

 An entirely different route towards realizing the proposed quantum (and classical) learning agents might employ condensed matter systems in which the proposed Markov chains could e.g.\ be realized through cooling or relaxation processes towards target distributions that then encode the state of belief of the agent. Here we envision rather non-trivial cooling/relaxation schemes in complex many-body systems, the study of which is also a prominent topic in the field of quantum simulation.  
 
In conclusion, it seems to us that the embodied approach to artificial intelligence acquires a further fundamental perspective by combining it with concepts from the field of quantum physics. The implications of embodiment are, in the first place, described by the laws of physics, which tell us not only about the constraints but also the ultimate possibilities of physical agents. In this paper we have shown an example of how the laws of quantum physics can be fruitfully employed in the design of future intelligent agents that will outperform their classical relatives in complex task environments. 
\\

\noindent\textbf{Acknowledgments:\\}
MAMD acknowledgs support by the Spanish MICINN grant FIS2009-10061, FIS2012-33152, the CAM research consortium QUITEMAD S2009-ESP-1594, the European Commission PICC: FP7 2007-2013, Grant No. 249958, and the UCM-BS grant GICC-910758. HJB acknowledges support by the Austrian Science Fund (FWF) through the SFB FoQuS: F 4012, and the Templeton World Charity Fund (TWCF) grant TWCF0078/AB46. 

GDP and VD have contributed equally to this work.

\section{Appendix}

\label{sect:Appendix}

\subsection{Formal definitions and behavior of reinforcement learning agents}
\label{subsect:Formaldefinitionsofreinforcementlearningagen2}
Here we formally define the model of reinforcement learning agents as employed in this work.
\DE (Reinforcement learning agent) \label{def:RLA}
A reinforcement learning agent is an ordered sextuplet $(\mathcal{S},\mathcal{A}, \Lambda,\mathcal{C}, \mathcal{D},\mathcal{U})$ where:
\begin{itemize}
\item $\mathcal{S} = \{s_1, \ldots, s_m\},  \mathcal{A} = \{a_1, \ldots, a_n\}$ are the sets of percepts and actions, respectively.
\item $\Lambda = \{0,1 \}$ is the set of rewards, offered by the environment.
\item $\mathcal{C} = \{C_1, \ldots, C_p\}$ is the set of possible internal states of the agent.
\item $\mathcal{D}: \mathcal{S} \times \mathcal{C} \rightarrow \mathcal{A}$ is the decision function, which outputs some action given a percept and the internal state.
\item $\mathcal{U}:  \mathcal{S} \times \mathcal{A} \times \Lambda \times \mathcal{C} \rightarrow \mathcal{C}$ is the update function, which updates the internal state based on the success or failure of the last percept-action sequence.
\end{itemize}
\ED

A few comments are in order.
In this work, the sets of percepts, actions and internal states are defined to be finite, but, in general, this need not be the case.
The set of rewards is binary, and this can again be generalized.
The update function may take additional information into account, based on additional outputs of the decision function, which are only processed internally, but this does not occur in the models we consider.

The decision function is not necessarily deterministic. In the non-deterministic case it can be formally defined as
\EQ{
\mathcal{D}: \mathcal{S} \times \mathcal{C} \rightarrow Distr(\mathcal{A})
}
that is, a function which takes values in the set of distributions over $\mathcal{A}$. In this case we also assume that this distribution is sampled before actual output is produced, and that the sampled action is the input to the update function.

Next, we consider equivalences between agents in so-called passive settings.

In the algorithmic tradition of machine learning, the learning pace is measured by external time (steps) alone, and the typical figure of merit is the percentage of rewarded actions of the agent, as the function of external time.
From an embodied agent perspective, this set-up corresponds to a special \emph{passive setting} where a static environment always waits for the responses of the agent.
This constraint imposes a restriction on the universality of statements which can at all be made about the performance of an agent. In particular, in that setting it is well-known that no two agents can be meaningfully compared without reference to a specific (or a class of) learning tasks - a collection of results dubbed `no free lunch theorems', and `almost no free lunch theorems'  \cite{1996_Wolpert,1997_Droste} \footnote{We acknowledge that the \emph{interpretation} of these results in the sense of their practical impact on the field are not without controversy. Nonetheless, the validity of the mathematical statements is not contested. See \cite{NFLorg} for more details.}. These results prove that when one agent outperforms another in a certain environment, there exists a different environment where the ranking according to performance is reversed. This, for instance, implies that every choice of environment settings, for which results of agent performance are presented, must be first well justified.
More critically for our agenda, in which we wish to make \emph{no} assumptions on the environment, passive settings would imply no comparative statements relating the performances of agents are possible.
In active scenarios internal time does matter, but nonetheless the passive setting plays a part. It provides a baseline for defining a \emph{passive} behavioral equivalence of agents, which will be instrumental in our analysis of active scenarios.

Let us denote the elapsed sequence of triplets $(percept, action, reward)$ which had occurred up to time-step $k$ (the \emph{history} of the agent) with $H_k$, for two agents $A$ and $A'$
who can perceive and produce the same sets of percepts ($\mathcal{S}$) and actions ($\mathcal{A}$), respectively.
Then we will say that $A$ and $A'$  are passively ($\epsilon-$)equal if at every external time step $k$ the probabilities $P_A$ and $P_{A'}$ of agents $A$ and $A'$, respectively, outputting some action $a \in \mathcal{A}$, given every percept $s \in \mathcal{S}$, and given all possible identical histories $H_k$ are  ($\epsilon-$)equal, in terms of the variational distance on distributions:
\EQ{
\dfrac{1}{2}\sum_{a} \vert P_{A}(a \vert H_k,s) - P_{A'}(a \vert H_k,s) \vert \leq \epsilon,
}
which we abbreviate with  \EQ{A \approx_{\epsilon} A'.}
If the agents considered are equipped with an extra parameter $\tau$ (a precision parameter), which fine tunes the behavior of the agent, we can demand more and require that the approximate equality above converges to an equality (i.e. $\epsilon \rightarrow 0$, as $\tau \rightarrow \infty$ ). Then in the limit, the relation above induces passive behavioral equivalence classes for fixed sets of possible percepts and actions.
In the case of the classical and quantum agents we consider in the main text, such precision parameters do exist, and, as we show later in this Supplementary Information, the approximate equality converges to an equality.

In passive settings, by definition, two passively equal agents perform equally well, and comparison of agents within a class is pointless. However, in the active scenario, with the classes in place, we can meaningfully compare agents within the same class, \emph{with no assumptions on the environment} \footnote{That is, a comparison can be made with no further assumptions on the environment beyond the trivial onesÑthat the percept and action sets are compatible with the environment and that the environment provides a rewarding scheme.}
Indeed, in an \emph{active} learning setting, two \emph{passively} equal agents $A$ and $A'$ may have vastly different success chances.
 To see this, suppose that the environment changes its policies on a timescale that is long compared to the internal timescale of agent $A$, but short relative to the internally slower agent $A'$.
 The best policy of both agents is to query the environment as frequently as possible, in order to learn the best possible actions. However, from the perspective of the slow agent, the environment will look fully inconsistent - once rewarded actions are no longer the right choice, as that agent simply \emph{did not have the time to learn}. Thus, in active scenarios, internal speed of the agent is vital.

\subsection{Classical and quantum walk basics}

\label{subsubsect:(Q)Wbasics}

A random walk on a graph is described by a MC, specified by a transition matrix $P$ which has entries $P_{ji} = \mathrm{Prob}( j | i )$. For an irreducible MC there exists a stationary distribution $\pi$ such that  $P\pi = \pi$. For an irreducible and aperiodic MC this distribution can be approximated by $P^t \pi_0$ i.e. by applying, to any initial distribution $\pi_0 $, the MC $t$ number of times where $t \ge t^{mix}_{\epsilon'} $.   This time is known as \emph{mixing time} and is defined as follows.
\DE (Mixing Time).

The mixing time is:
\EQ{
t^{mix}_{\epsilon'} = \min \{ t | \forall s \ge t, \pi_0 : || P^s \pi_0 - \pi || \le \epsilon' \}\nonumber
}
\ED
The latter can be related to the spectral properties of the MC P, in the case of reversible chains, via the following theorem \cite{1993_Sinclair}:
\TH
\label{mixing}
The mixing time satisfies the following inequalities:
\EQ{
\dfrac{1}{\delta} \dfrac{\lambda_2}{ \log 2 \epsilon'} \le  t^{mix}_{\epsilon'} \le \dfrac{1}{\delta} (\max \log \pi_i^{-1} + \log (\epsilon')^{-1})\nonumber
}
\HT
Here, we use $\epsilon'$ instead of the standard $\epsilon$ for consistency with the rest of the Supplementary Information, as $\epsilon$ has a reserved meaning.

For the purpose of clarity let us introduce some definitions and theorems, originally provided in \cite{2004_Szegedy_IEEE,2011_Magniez_SIAM} that will be useful to introduce the notation and to prove the main results for the speedup of quantum agents.

The quantum analog of the applying the MC P is given by:
\DE (Quantum Diffusion Operators).

The quantum diffusion operators, the analogs of the classical diffusion operators are given by the the following transformations:
\EQ{
 U_P: \ket{i}\ket{0} \mapsto \ket{i} \ket{p_i}
 }
 \EQ{
V_P: \ket{0}\ket{j} \mapsto  \ket{p^*_j} \ket{j}\, ,
}
where $\ket{p_i} =  \sum_j \sqrt{P_{ji}}\ket{j}$, $\ket{p^*_j} =\sum_i \sqrt{P^*_{ij}}\ket{i}$ and $P^*$ is the time reversed MC  defined by $\pi_i P_{ji} = \pi_j P^*_{ij}$.
\ED%
We will consider the application of the MC P (for the classical agent) and the quantum diffusion operators (for the quantum agent) as the (equally time consuming) primitive processes, as is done in the theory of quantum random walks \cite{2011_Magniez_SIAM}.
Next, we can define the quantum walk operator $W(P),$ for the Markov chain P.

\DE (Walk Operator or Quantum Markov Chain).

The walk operator or Quantum Markov Chain is  given by
\EQ{
W(P) = (2\Pi_2- \one) (2\Pi_1- \one),
}
where $\Pi_1$ is the projection operator onto the space $\mathrm{Span}\{\ket{i} \ket{p_i}\}_i$ and $\Pi_2$ is the projection operator onto
$ \mathrm{Span} \{\ket{p^*_j} \ket{j} \}_j$.
\ED

The quantum walk operator can be easily realized through four applications of the quantum diffusion operators, see e.g. \cite{2011_Magniez_SIAM} for details. 

Another standard operation which both the classical and the quantum agents do is checking whether the clip found is flagged (corresponding to checking whether an item is marked). The quantum check operator is defined as follows.
\DE (Check).
The quantum check operator is the reflection denoted as $\mathrm{ref}(f(s))$ performing
\EQ{
\ket{i}\ket{j} \mapsto \left \lbrace  {-\ket{i}\ket{j} ,\ \textup{if} \ i\in f(s)  \atop \ket{i}\ket{j} ,\ \textup{otherwise}} \right.,
}
where $f(s)$ denotes the set of flagged actions corresponding to the percept $s$.
\ED In order to prove our main theorems we will be using the ideas introduced in the context of quantum searching \cite{2004_Szegedy_IEEE,2011_Magniez_SIAM} which we now briefly expose. In the quantum walk over graphs approach to searching, one defines an initial state, which encodes the stationary distribution of a MC,
\EQ{
\ket{\pi } = \sum_i \sqrt{\pi(i)}\ket{i} \ket{p_i} ,
}
and performs a rotation onto the state containing the 'marked items'
\EQ{
\ket{\tilde{\pi}} :=  \dfrac{\Pi^{f(s)} \ket{\pi }}{|| \Pi^{f(s)}  \ket{\pi} ||}, }
where $\Pi^{f(s)} $ is the projector on the space of marked items i.e $\Pi^{f(s)} = \sum_{i \in f(s)} \ket{i}\bra{i}\otimes \one .$ Let us point out that $\textrm{Span} \{\ket{\tilde{\pi}}, \ket{\pi} \} \subseteq \mathrm{Span}\{\ket{i} \ket{p_i}\}_i +  \mathrm{Span} \{\ket{p^*_j} \ket{j} \}_j$.

In order to achieve this rotation one makes use of two reflections. The first is the reflection over $\ket{\tilde{\pi}^\bot}$ (denoted $\mathrm{ref} (\ket{\tilde{\pi}^\bot}) $ ), the state orthogonal to $\ket{\tilde{\pi}}$ in $\textrm{Span} \{\ket{\tilde{\pi}}, \ket{\pi} \}$. This operator can be realized using the primitive of checking. Indeed, we have the following claim (stated in \cite{2011_Magniez_SIAM}) given by:
\LE \label{mulemma}
Restricted on the subspace $\mathrm{Span} \{\ket{\tilde{\pi}}, \ket{\pi} \}$, the action of $\mathrm{ref} (\ket{\tilde{\pi}^\bot}) $ is identical to $ - \mathrm{ref} (f(s))$.
\EL
\proof Let $ \alpha \ket{\pi} + \beta \ket{\tilde{\pi}}$ be a vector in $\mathrm{Span} \{\ket{\tilde{\pi}}, \ket{\pi} \}$. We have that:
\EQ{
- \mathrm{ref}(f(s)) \left( \alpha \ket{\pi} + \beta \ket{\tilde{\pi}}\right) =
}
$$
= \alpha (\one - 2 \Pi^{f(s)}) \ket{\pi} - \beta \ket{\tilde{\pi}}
$$
where $\Pi^{f(s)} $ is the projector on the set $f(s) $.
The result easily follows by noting that
\EQ{
\mathrm{ref}(\ket{\tilde{\pi}^\bot}) \left( \alpha \ket{\pi} + \beta \ket{\tilde{\pi}}\right) =  }
$$ = \alpha \left(\one - 2\dfrac{\Pi^{f(s)} \ket{\pi}\bra{ \pi} \Pi^{f(s)}}{|| \Pi^{f(s)} \ket{\pi} ||^2}\right) \ket{\pi} - \beta \ket{\tilde{\pi}},$$
and that
\EQ{
\dfrac{\Pi^{f(s)} \ket{\pi}\bra{ \pi} \Pi^{f(s)}}{|| \Pi^{f(s)} \ket{\pi} ||^2} \ket{\pi} = \dfrac{\bra{\pi} \Pi^{f(s)} \ket{\pi}}{|| \Pi^{f(s)} \ket{\pi} ||^2}\Pi^{f(s)} \ket{\pi}  \nonumber }
$$= \Pi^{f(s)} \ket{\pi}, \nonumber$$
 since 
 $\dfrac{\bra{\pi} \Pi^{f(s)} \ket{\pi}}{|| \Pi^{f(s)} \ket{\pi} ||^2} = 1$.

\qedhere

On the other hand, the reflection over $\ket{\pi}$ is not straightforward.
One can devise an approximated scheme to implement this reflection using the phase estimation algorithm. Indeed, one can build a unitary operator, using phase estimation applied to the quantum walk operators, which approximates the reflection over $\ket{\pi}$.
Before we state the theorem regarding this approximate reflection operator (constructively proven in \cite{2011_Magniez_SIAM}), we will first give another result regarding the spectrum of the quantum walk operator, which will be relevant to us presently.
\TH \label{Wspectral} (Szegedy \cite{2004_Szegedy_IEEE})
Let P be an irreducible, reversible MC with stationary distribution $\pi$. Then the quantum walk operator $W(P)$ is such that:
\begin{enumerate}
\item[1.] $W(P) \ket{\pi} = \ket{\pi}$.
\item[2.] $W(P) \ket{\psi} = \exp (\pm 2i \theta )\ket{\psi} $,  where $ \cos (\theta) = | \lambda|$ is the absolute value of an eigenvalue of $P$ and $\ket{\psi} \in \mathrm{Span}\{\ket{i} \ket{p_i}\}_i +  \mathrm{Span} \{\ket{p^*_j} \ket{j} \}_j$.
\item[3.] $W(P)$ has no other eigenvalue in $\mathrm{Span}\{\ket{i} \ket{p_i}\}_i +  \mathrm{Span} \{\ket{p^*_j} \ket{j} \}_j$.
\end{enumerate}
\HT
Note that the  phase gap $\Delta$, defined as the minimum nonzero $2 \theta $, is such that $\cos \Delta = |\lambda_2|$, where $\lambda_2$ is the second-largest eigenvalue of $P$ with respect to the absolute value. One can then, with some algebra, conclude that $\Delta \ge 2 \sqrt{\delta}$.

Let us note that any unitary able to approximately detect whether the eigenvalue of $W(P)$ of a state in $\mathrm{Span}\{\ket{i} \ket{p_i}\}_i +  \mathrm{Span} \{\ket{p^*_j} \ket{j} \}_j$ is different from one (or equivalently, its eigenphase is different from zero) and conditionally flip the state, will do. We will use such a unitary to approximate  $ \mathrm{ref}(\ket{\pi})$.
Let us use this intuition to build such a unitary, $R(P)$, that takes as a parameter the precision $s$ and refer to it in the following as the approximate reflection operator $R(P)$~\cite{2011_Magniez_SIAM}:
\TH  \label{def:ARO} (Approximate Reflection Operator \cite{2011_Magniez_SIAM}).

Let P be an ergodic, irreducible Markov chain on a space of size $|X|$ with (unique) stationary distribution $\pi$. Let $W(P)$ be the corresponding quantum Markov Chain with phase gap $\Delta$.
Then, if $s$ is chosen in $ \mathcal{O}  ( \log_2 (1 / \Delta ) ) $ , for every integer $k$ there exist a unitary $R(P)$ that acts on $2 \lceil  \log_2 |X| \rceil + ks $ qubits, such that:
\begin{enumerate}
\item[1.] $R(P)$ makes at most $k 2^{s+1} $ calls to the (controlled) $ W(P)$ and $ W(P)^\dagger $.
\item[2.] $R(P) \ket{\pi}\ket{0}^{\otimes ks} = \ket{\pi}\ket{0}^{\otimes ks} $ .
\item[3.]  If $ \ket{\psi} \in \mathrm{Span}\{\ket{i} \ket{p_i}\}_i + \mathrm{Span} \{\ket{p^*_j} \ket{j} \}_j$and is orthogonal to $\ket{\pi}$, then $ ||\left(R(P) + \one \right)\ket{\psi}\ket{0}^{\otimes ks} || \le 2^{1-k} $.
\end{enumerate}
\HT
By the approximate reflection theorem above, there exists a subroutine $R(P)(q,k)$, where $s$ from the statement of Theorem~\ref{def:ARO}  is taken as $log_2(q)$, and $k$ explicitly controls the fidelity of the reflection. Note that, in the definition of the quantum reflecting agent from the main text, the parameter $q$ was chosen in  $\tilde{O}(1/\sqrt{\delta}),$ and since by a Theorem of Szegedy \cite{2004_Szegedy_IEEE}, as we have commented, it holds that $1/\sqrt{\delta} \in O(1/\Delta)$, we have that the fidelity of the approximation reflection approaches unity exponentially quickly in $k$.
We note that the parameter $k$ should be additionally increased by a logarithmic factor of $O(log (1/\sqrt{\epsilon}))$, in order to compensate for the accumulated error stemming from the iterations of the ARO operator, which, as clarified, we omit in this analysis.

For the explicit construction of the approximate reflection operators, we refer the reader to \cite{2011_Magniez_SIAM}.

\subsection{The PS model}
\label{abst:ThePSModel}

The PS model is a reinforcement learning agent model, thus it formally fits within the specification provided with Def. \ref{def:RLA}. Here we will recap the standard PS model introduced in \cite{2012_Briegel} but note that the philosophy of the projective simulation-based agents is not firmly confined to the formal setting we provide here, as it is more general.
PS agents are defined on a more conceptual level as agents whose internal states represent episodic and compositional memory and whose deliberation comprises an association driven hops between memory sequences - so called \emph{clips}.
Nonetheless, the formal definitions we give here allow us to precisely state our main claims.
Following the basic definitions, we provide a formal treatment of a slight generalization of the standard PS model which subsumes both the standard and the reflecting agent model we refer to in the main text, and formally treat in this Supplementary Information later.

The PS model comprises the percept and action spaces as given in Def.~\ref{def:RLA}.
The central component of PS is the so-called episodic and compositional memory (ECM), and it comprises the internal states of the agent.
The ECM is a directed weighted network (formally represented as a directed weighted graph) the vertices of which are called \emph{clips}.

Each clip $c$ represents fragments of episodic experiences, which are formally tuples
\EQ{
c = (c^{(1)}, c^{(2)}, \ldots, c^{(L)}),
}

where each $c_k$ is \emph{an internal representation} of a percept or an action, so
\EQ{
c^{(k)} = \mu(i), \ i\in \mathcal{S} \cup \mathcal{A},
}
where $\mu$ is a mapping from \emph{real} percepts and actions to the \emph{internal representations}.
We will assume that each ECM always contains all the unit-length clips denoting elementary percepts and actions.

Within the ECM, each edge between two clips $c_i$ and $c_j$ is assigned a weight $h(c_i, c_j) \geq 1$ and the weights are collected in the so-called $h-$matrix. The elementary process of the PS agent is a Markov chain, in which the excitations of the ECM hop from one clip to another, where the transition probabilities are defined by the $h$-matrix:
\EQ{
p(c_j \vert c_i) = \dfrac{h(c_i, c_j) }{\sum_k h(c_i, c_k) },
}
thus the $h-$matrix is just the non-normalized transition matrix.
In the standard PS model, the decision function is realized as follows: given a percept $s$, the corresponding clip in the ECM is excited and hopping according to the ECM network is commenced.
In the simplest case, the hopping process is terminated once a unit-length action clip is encountered, and this action is coupled out and output by the actuator (see Fig. \ref{fig:AgentBasic}). The moment when an action is coupled out can be defined in a more involved way, as we explain presently.

Finally, the update rule, in the standard model necessarily involves the re-definition of the weights in the $h$-matrix.
A prototypical update rule, for a fully classical agent, defining an update from external time-step $t$ to $t+1$ depends on whether an action has been rewarded. If the previous action has been rewarded, and the transition between clips $c_i,c_j$ had actually occurred in the hopping process then
the update is as follows:
\begin{equation}
\begin{split}
h^{(t+1)}(c_i, c_j)=\\ h^{(t)}(c_i, c_j) - \gamma(h^{(t)}(c_i, c_j) -1) + \lambda
\end{split}
\end{equation}
where $0<\lambda$ is a positive reward and $0\leq\gamma\leq1$ is a dissipation (forgetfulness) parameter.
If the action had not been rewarded, or the clips $c_i,c_j$ had not played a part in the hopping process then the weights are updated as follows:
\EQ{
h^{(t+1)}(c_i, c_j)=h^{(t)}(c_i, c_j) - \gamma(h^{(t)}(c_i, c_j) -1).
}

The update rule can also be defined such that the update only requires the initial and terminal clip of the hopping process, which is always the case in the \emph{simple} PS model, where all the clips are just actions or percepts, and hopping always involves a transition from a percept to an action. This example was used in section \ref{subsect:comparison}. For that particular example, the update function can be exactly defined by the rules above.
As mentioned, aside from the basic structural and diffusion rules, the PS model allows for additional structures, which we repeat here. 1) \emph{Emoticons} - the agents short term memory, i.e. \emph{flags} which notify the agent whether the currently found action, given a percept was previously rewarded or not. For our purposes, we shall use only the very rudimentary mode of flags, which designate that the particular action (given a particular percept) was not already \emph{un}successfully tried before. If it was, the agent can 'reflect on its decision' and re-evaluate its strategies, by re-starting the diffusion process.
This reflection process is an example of a more complicated out-coupling rule we have mentioned previously.
2) \emph{Edge and clip glow} - mechanisms which allow for the establishing of additional \emph{temporal correlations}.
3) \emph{Clip composition} - the PS model based on episodic and compositional memory allows the creation of new clips under certain variational and compositional principles. These principles allow the agent to develop \emph{new behavioral patterns} under certain conditions, and allow for a dynamic reconfiguration of the agent itself. For more details we refer the reader to \cite{2012_Briegel, Julian12,Briegel2_2012}.

As illustrated, the PS model allows for a great flexibility. A straightforward generalization would allow for the ECM network to be percept-specific, which is the view we adopt in the definition of reflecting agents.
However, the same notion can be formalized without introducing multiple networks (one for every percept). In particular, the ECM network allows for action and percept clips to occur a multiple number of times.
Thus the ECM network can be represented as $|\mathcal{S}|$ disjoint sub-networks, each of which comprises all elementary action clips, and only one elementary percept clip. This structure is clearly within the standard PS model, and it captures all the features of the reflecting PS agent model. A simple case of such a network, relative to the standard picture, is illustrated in Fig. \ref{fig:Comparison}, part b).
Thus, the reflecting PS agent model is, structurally, a standard PS model as well.

\subsection{Behavioral equivalence of classical and quantum reflecting agents}
\label{subsect:Formaldefinitionsofreflectingagents}

In this section, we show that the classical and quantum reflecting agents (r-PS), denoted $A_{r-PS}^c$ and $A_{r-PS}^q$ are approximately equal, that is $A_{r-PS}^c \approx_{\alpha} A_{r-PS}^q,$ where $\alpha$ can be made arbitrarily small without incurring a significant overhead in the internal time of the agents.

We do so by separately showing that the output distributions of both the classical and the quantum r-PS are  $\alpha$-close to the previously mentioned tailed distribution, for an arbitrarily small $\alpha$. The main claim will then follow by the triangle inequality on the behavioral distance (which holds since the behavioral distance is the variational distance on the output distributions).

For completeness, we begin by explicitly giving the deliberation process of the classical r-PS, described in the main text, and proceed with the behavioral theorem for classical r-PS.

The agent's decision-making process (implementing the decision function $\mathcal{D}$, see section \ref{subsect:Formaldefinitionsofreinforcementlearningagen2} of this Supplementary Information for details), given percept $s$, is given by the following steps.

Let $t_{mix}^c \in \tilde{O}( 1/\delta_{s})$;
\begin{enumerate}
\item Sample $y$ from some fixed distribution $\pi_0$.
\item Repeat:\\
\begin{enumerate}
\item \textbf{Diffuse:} (re-)mix the Markov chain by $$\pi = P_s^{t_{mix}^c} y.$$
\item  \textbf{Check:} Sample $y$ from $\pi$. If $y$ is a flagged action, break and output $y$.
\end{enumerate}
\end{enumerate}

\label{subsect:twomain}

In the following, when $\pi$ and $\pi'$ are distributions then $\|\pi - \pi' \|$ denotes the standard variational distance (Kolmogorov distance) on distributions, so $$\|\pi - \pi' \| = \dfrac{1}{2} \sum_{x} |\pi(x) - \pi'(x)  |.$$

\TH (Behavior of classical reflecting agents)

Let $P_s$ be the transition matrix of the Markov chain associated to percept $s$, let $f(s)$ be the (non-empty) set of flagged action clips. Furthermore, let $\pi_s(x)$ be the probability mass function of the stationary distribution $\pi_s$ of $P_s$ and let $\tilde{\pi}_s$ be the renormalized distribution of $\pi_s,$ where the support is retained only over the flagged actions, so:
\EQ{
\tilde{\pi}_s(i) = \left \lbrace  {\dfrac{\pi_s(i)}{\sum_{j \in f(s)} \pi_s(j)},\ \textup{if} \ i\in f(s)  \atop 0, \textup{otherwise}} \right. \label{tailed}}
Let $\kappa$ be the probability distribution over the clips as outputted by the classical reflecting agent, upon receiving $s$. Then the distance
$\| \kappa - \tilde{\pi}_s \|$ is constant (ignoring logarithmic factors), and can be efficiently made arbitrarily small.
\HT

\proof

Note that, since the Markov chain is regular, by Theorem \ref{mixing}, the distribution $P_s^{\tilde{O}(1/\delta_s)} \pi_0$, for any initial distribution $\pi_0$ is arbitrarily close to the stationary distribution of $P_s$, more precisely,
\EQ{
\| P_s^{(k_0 + \log (1/\epsilon'))/\delta_s)} \pi_0 - \pi\| \leq  \epsilon',
}
or,  equivalently
\EQ{
\| P_s^{(k_0 + k_1)/\delta_s)} \pi_0 - \pi\| \leq  e^{-k_1},
}
where $k_0 = \max_{i} \log(\pi(i)^{-1}).$
While $\pi(i)^{-1}$ can in principle be very large, it only logarithmically contributes to the overhead, so it can be effectively bounded, and omitted from the analysis.
Thus, we can achieve an exponentially good approximation of the stationary distribution with $\tilde{O}(1/\delta_s)$ iterations. In the remainder we will denote $\pi' =P_s^{\tilde{O}(1/\delta_s)} \pi_0$

The reflecting agent mixes its Markov chain (achieving $\pi'$) and then samples from this distribution, iteratively until a flagged action is hit. 
Thus $\kappa = \tilde{\pi}'$, where  $\tilde{\pi}'$ is the tailed $\pi'$ distribution in the sense of Eq. (\ref{tailed}) (substituting $\pi_s$ with $\pi'$).

Hence we have that 
\EQ{
\begin{split}
\| \kappa - \tilde{\pi}_s \| = \| \tilde{\pi}' - \tilde{\pi}_s  \|.
\end{split}
}

Next, we only need to bound  $\| \tilde{\pi}' - \tilde{\pi}_s  \|$. 
Note that 
\EQ{
\| \tilde{\pi}' - \tilde{\pi}_s  \| = \| \dfrac{1}{\epsilon_{\pi'}} \pi'_\textup{sub} -  \dfrac{1}{\epsilon} {(\pi_s)}_\textup{sub} \|
}
where $\epsilon$ and $\epsilon_{\pi'}$ are the probabilities of sampling a flagged action from $\pi_s$ and $\pi'$, respectively, and   $\pi'_\textup{sub}$ and ${(\pi_s)}_\textup{sub}$ are the sub-normalized distributions  $\pi'_\textup{sub} = \epsilon_{\pi'} \  \tilde{\pi}'$ 
and ${(\pi_s)}_\textup{sub} = \epsilon \  \tilde{\pi}_s$.
Note that it holds that $\| \pi'_\textup{sub} - {(\pi_s)}_\textup{sub} \| \leq \| \pi' - \pi_s\|$.
So we have that
\EQ{
\begin{split}
\| \dfrac{1}{\epsilon_{\pi'}} \pi'_\textup{sub} -  \dfrac{1}{\epsilon} {(\pi_s)}_\textup{sub} \| = \\
 \| \dfrac{1}{\epsilon_{\pi'}} \pi'_\textup{sub} -\dfrac{1}{\epsilon} \pi'_\textup{sub}+\dfrac{1}{\epsilon}\pi'_\textup{sub} -  \dfrac{1}{\epsilon} {(\pi_s)}_\textup{sub} \| \leq\\
| \dfrac{1}{\epsilon_{\pi'}} -  \dfrac{1}{\epsilon}| \|  \pi'_\textup{sub}  \| + \dfrac{1}{\epsilon} \|\pi'_\textup{sub} - {(\pi_s)}_\textup{sub} \| \leq\\
| \dfrac{1}{\epsilon_{\pi'}} -  \dfrac{1}{\epsilon}| \epsilon_{\pi'} +  \dfrac{1}{\epsilon} \|\pi' - \pi_s\| \leq \dfrac{|\epsilon - \epsilon_{\pi'}|}{\epsilon} + \dfrac{e^{-k_1}}{\epsilon}.
\end{split}
}
Next note that $ e^{-k_1} \geq \| \pi' - \pi_s \| \geq |  \epsilon_{\pi'} - \epsilon| $, so finally we have
\EQ{
\| \tilde{\pi}' - \tilde{\pi}_s  \| \leq 2 \dfrac{e^{-k_1}}{\epsilon} = 2 e^{-k_1 + log(1/\epsilon)}.
}
Since we are interested in the analysis at the $\tilde{O}$ level, we can omit the logarithmically contributing factor above.
This concludes the proof \qed.

Note that since the ideal stationary distribution $\pi$ (of MC $P_s$) and the approximation realized by mixing are no further than $e^{-k_1}$ apart, and since the expected time of producing an action when sampling from the ideal distribution is $O(1/\epsilon_s),$ this entails that the expected number of checks the classical r-PS will perform is in $\tilde{O}(1/\epsilon_s)$.

Next, we prove an analogous theorem for the quantum agents.

 \TH (quantum reflecting agents)
Let $P_s$ be the transition matrix of the Markov chain associated to percept $s$, let $f(s)$ be the (non-empty) set of flagged action clips. Furthermore, let $\pi_s(i)$ be the probability mass function of the stationary distribution $\pi_s$ of $P_s$ and let $\tilde{\pi}_s$ be the renormalized distribution of $\pi_s,$ where the support is retained only over the flagged actions, so:
\EQ{
\tilde{\pi}_s(i) = \left \lbrace  {\dfrac{\pi_s(i)}{\sum_{j \in f(s)} \pi_s(j)},\ \textup{if} \ i\in f(s)  \atop 0, \textup{otherwise}} \right.}
Let $\kappa$ be the probability distribution over the clips as output by the quantum reflecting agent, upon receiving $s$. Then the distance
$ \| \kappa - \tilde{\pi} \|  $
 is constant (up to logarithmically contributing terms), and can be efficiently made arbitrarily small.
\HT

\proof

The proof consists of two parts. First, we prove that, if the reflection operators (used in the diffusion step of the quantum agent) are ideal, then the claim follows. The remainder of the proof, which considers imperfect reflection operators which are actually used, follows from the proof of Theorem 7 in \cite{2007_Magniez_SIAM}.
In the following, we will, by abuse of notation, denote with $\ket{\pi_s}$ both the coherent encoding of the stationary distribution, and the state $\ket{\pi_{init}},$ which is the initial state of the quantum agent, and also the state over which reflections occur, following the notation of \cite{2007_Magniez_SIAM}.  Note that the latter state is obtained by applying the quantum diffusion operator $U_P$ once to $\ket{\pi_s} \ket{0}$, and the statistics of measuring the first register in $\ket{\pi_{init}}$ match the statistics of measuring $\ket{\pi_s}$ in the computational basis.

 Assuming the procedure the agent follows starts from the perfect stationary distribution, and that the reflections over $\ket{\pi_s}$ are perfect then, by Lemma \ref{mulemma} we have that the state of the system never leaves the span of $\ket{\tilde{\pi}_s}$ and $\ket{\tilde{\pi}_s^\bot}$ where  $\ket{\tilde{\pi}_s^\bot}$ is the state orthogonal to  $\ket{\tilde{\pi}_s}$  in $\mathrm{Span}\{\ket{\tilde{\pi}_s}, \ket{\pi_s} \}$.

Note that the agent outputs an action only conditional on the result being a flagged action. This is equivalent to first projecting the state of the system onto the subspace of flagged actions, using the projector $\Pi^{f(s)} = \sum_{x\in f(s)} \dm{x}$, followed by renormalization and a measurement. The state after this projection is, for any state in $\mathrm{Span}\{\ket{\tilde{\pi}_s}, \ket{\pi_s} \}$ a (sub-)normalized state $\ket{\tilde{\pi}_s},$ hence, after normalization, measurement outcomes always follow the distribution $\tilde{\pi}_s$.

However, any state in $\mathrm{Span}\{\ket{\tilde{\pi}_s}, \ket{\pi_s} \}$ which is not exactly $\ket{\tilde{\pi}_s}$ still has a non-zero support on the non-flagged clips. However, it is the key feature of Grover-like search algorithms (like the search algorithm in \cite{2007_Magniez_SIAM}, the reflections of which make up the decision process of the quantum agent) that the reflection iterations 
produce a state which has a constant overlap with the target state (in our case $\ket{\tilde{\pi}_s}$, which has support only over flagged actions).
This property implies that the probability of failing to hit a flagged action is at most some constant $\beta<1$. Now, if the deliberation process  is iterated some constant $k_3$ number of times it is straightforward to see that 
  $ \| \kappa - \tilde{\pi}_s \| \leq \beta^{k_3},  $ thus decays exponentially quickly as desired.
  
Next, we need to consider the errors induced by the approximate reflections. 

The analysis in \cite{2007_Magniez_SIAM} (proof of Theorem 7) directly shows that the terminal quantum (after $t_2' \in [0, t_2]$ iterations) state $ \ket{\psi} $ is close to the state $\ket{\phi}$ the algorithm would produce had the reflections been perfect, formally,
\EQ{
\|\ket{\psi} - \ket{\phi} \| \leq 2^{1-c},
}
where $c$ is a constant only additively increasing the internal parameter $k$ of the approximate reflection subroutine. 

Thus the error on the final state produced by the quantum agent induced by the approximate reflection algorithm can be made arbitrarily small within the allowed cost $\tilde{O}(1/\sqrt{\delta_s}),$
which also means that all distributions obtained by measurements of these states will not differ by more than $4 \times2^{1-c}$ \cite{1993_Vazirani}.

Returning to the inequality proven for the perfect reflections, we see that the error of magnitude $4 \times2^{1-c}$ can only increase the probability of failing to output a flagged action from $\beta$ to $4 \times2^{1-c} + \beta$. Thus, by tuning the parameters only linearly, we can make sure the agent produces an action with an exponentially high probability in terms of the parameter $k_3$ (the number of iterations of the deliberation process).
If an action has been produced, then, as we have shown above, it holds that it has been sampled from a distribution within $4 \times2^{1-c}$ distance of the tailed distribution $\tilde{\pi}_s$, which again can be efficiently made exponentially small. Thus the theorem holds. \qed

A corollary of the two theorems above is that the classical and quantum agents are passively approximately equal.
But then, from the definitions of these embodied agents we can see that the quantum agent exhibits a consistent quadratic speed up, in terms of the required number of applications of their elementary operations. That is, the quantum agent is quadratically faster as claimed. This proves the main claim of this paper.

The quantum agent may be required to reiterate the deliberation procedure if the final measurement finds a non-action clip. This procedure is easy if multiple copies of the initial state are always available. Alternatively, a way of re-creating the desired initial coherent encoding of the stationary distribution can be achieved by `inverting' the quantum search algorithm in \cite{2011_Magniez_SIAM}  by `unsearching'  the found non-action clip -- that is, by inverting the (unitary) Grover iterations which would be applied to in order to search for the found clip from the initial state. This process will, with high fidelity, again recreate a good approximation of the initial stationary distribution \footnote{Additionally, the quality of the approximation can be, probabilistically, arbitrarily raised by one round of phase estimation algorithm for the walk operator, provided no phase is detected. Should this fail, the `unsearching' process is repeated. }, but may be costly if the found non-action clip has a low frequency in the stationary distribution. 
A more efficient resolution to the problem of reiteration of deliberation can be achieved if the quantum agent can, additionally, perform a generalized measurement given by a positive operator-valued measure (POVM), which projects the state produced by the Grover-like iterations to the subspaces spanned by non-flagged and flagged clips, defined by POVM elements
\EQ{
\Pi^{f(s)} = \sum_{i\in f(s)} \ket{i} \bra{i},\\
\Pi^{\textup{non}-f(s)} = \mathbbmss{1} - \Pi^{f(s)}.
}
If the first outcome of this measurement is obtained, the agent will output the required action according to the desired distribution, by measuring the residual state.
If the other outcome is obtained, then the resulting state is the coherent encoding of the stationary distribution, re-normalized so that it has support only over non-flagged clips. Starting from this state (which is in the work space of the Grover-like iterations), the agent can `unsearch' for all non-flagged clips to regain the initial distribution.
Crucially, the number of steps here will scale as $\tilde{O}(1/\sqrt{1-\epsilon_s})$ which will in interesting cases be effectively constant.

\subsection{Comparison of PS agent models}
\label{subsect:comparison}

In this section we compare reflecting PS agents with the standard PS agents, which have been previously studied \cite{2012_Briegel, Julian12}. The main difference between the standard and the reflecting model is that the standard agents evolve their Markov chain until the first instance where an action clip has been hit, whereas the reflecting agents allow for the Markov chain to fully mix.
In the standard PS model, the underlying Markov chain should never mix, as indeed in that case, the construction will guarantee that the actual action performed is independent from the received percept (if the Markov chain is irreducible).
This suggests that large clip networks of high conductivity which could, e.g., appear in agents which have undergone a rich complex set of episodic experiences (for instance any type of clip compositions we mentioned previously) are better suited for the reflecting, rather than the standard PS agents.

Conversely, since the reflecting agent must always fully mix its Markov chain, and the standard does not, it would seem that the standard model should typically outperform the reflecting model on simple clip networks (as the hitting time may be much smaller than the mixing time).
 This would suggest that the standard and reflecting PS models should not be compared on simple structures, as no quantum advantage can be demonstrated.

 In contrast to these hypotheses, here we show that even for the simplest (but non-trivial) standard PS agent construction \cite{2012_Briegel,Julian12}, there exists a natural reflecting agent analog, where the performance of the classical reflecting agent matches the standard PS construction, and consequently, the quantum reflecting agent yields a quadratic speed-up over both.

 \subsubsection{Simple standard PS agent with flags}

Recall that, in the standard PS model, the internal state of the agent is defined by an $h-$matrix, which defines the transition probabilities of the directed Markov chain over the clip space.
In the simplest case, the clip space comprises only percepts and actions, the agent is initialized so that each percept is connected to each action with unit weight (implying equiprobable transitions), and no other connections exist.
Furthermore, the update function we assume is the standard update function as defined in \cite{2012_Briegel} and repeated in the Appendix \ref{abst:ThePSModel}. In this particular model, the internal elementary time-step is one evolution of the Markov chain, so the agent always decides in one step, and there is nothing to improve.
However, if we introduce short term memory, which significantly improves the performance of the model \cite{2012_Briegel}, then the fact that the agent may hit an un-flagged item implies it may have to walk again. If $\epsilon_s$ denotes the probability (for the clip $s$) that the agent hits a flagged item, then the expected number of elementary transitions, and checks, the agent will perform is $O(1/\epsilon_s)$ (see Fig.~\ref{fig:Comparison} for an illustration).
For this model, we next introduce a direct reflecting agent analog.

\begin{figure}
\label{fig:Comparison}
\includegraphics[trim=2cm 0cm 5cm 0cm,width=0.5\textwidth]{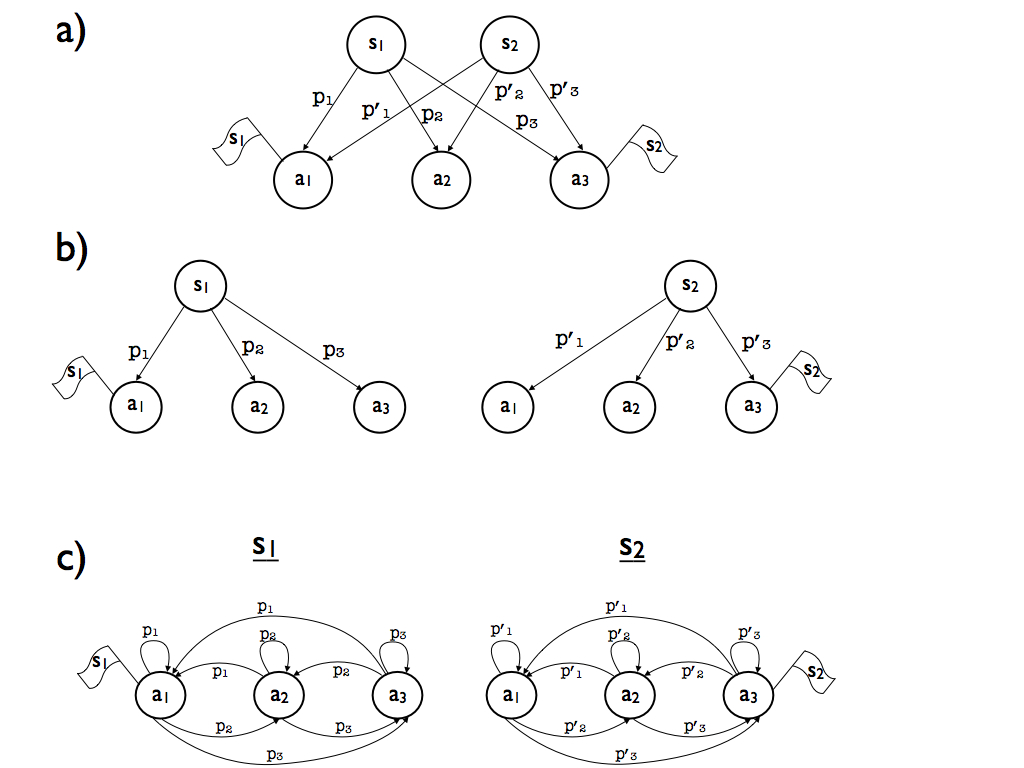}
\caption{
a) The simple PS model with flags. Transition from percept to action occurs in one step. Flags are percept-dependent.
b) Intermediary step: The graph of a) can be broken up into two graphs, by duplication of actions.
c) The straightforward analog of the simple PS in the reflecting agent model. To each percept, a MC over the actions only is assigned, such that the stationary distribution recovers the initial output probabilities. Flags are trivially inherited.
}
\end{figure}

\subsubsection{Simple reflecting agent with flags}

%
%

It is relatively straightforward to construct a reflecting PS agent which is behaviorally equivalent to the simple standard agent with flags, given above.
To each percept we assign a Markov chain over just the action space, such that the transition matrix is column-constant, and the column simply contains the transition probabilities for that particular percept in the standard model above, see Fig.~\ref{fig:Comparison} for an illustration.
The stationary distribution of this Markov chain then matches the desired transition probabilities.
The update function for this agent is then exactly the same as in the standard model (that is, internally, the reflecting agent can also keep and update the same $h-$matrix which induces the Markov chains), and the flags are also treated in exactly the same way.

Note that the transition matrix of such a Markov chain is rank-1, meaning it has only one non-zero eigenvalue, and since the trace of this matrix is unit (and trace is invariant under diagonalization/basis change), this implies the largest eigenvalue is 1 and all the others are zero.
Thus we have that the spectral gap is always $1$.
This spectral gap immediately implies that the Markov chain is fully mixed after just a single transition (as $\delta=1$).
Thus the classical reflecting agent will also perform $O(1/\epsilon_s)$ transitions and checks until an action is performed, given the percept $s$.
However, the quantum agent only requires $O(1/\sqrt{\epsilon_s})$ calls to the quantum walk operator.
Thus, even in the very simple setting we see that the reflecting agents can be compared to the standard PS agent model, and maintain a generic quadratic speed-up when quantized.

This simple model also serves as an illustration of a setting in which the initial state $\ket{\pi_s}$ is easily prepared.
Since the relevant transition matrix $P_s$ is of rank-1 it is straightforward to see that one application of the quantum diffusion operator $U_{P_s}$ applied on the register initialized to $\ket{1} \ket{0}$ generates the state $\ket{1} \otimes \sum_{j} \sqrt{P_{j,1}} \ket{j} = \ket{1}\ket{\pi_s}$,  since the $P_s$ is a column-constant matrix containing the stationary distribution as columns. 

While the scenario considered in this section is somewhat restricted, it is not without importance as for the standard PS agents we already have a body of results in the classical case \cite{2012_Briegel, Julian12}. 
We emphasize, however, that the quadratic speed-up proven in this work is not restricted to this scenario. 

\label{sect:ComparisonofPSagents}

\bibliographystyle{unsrt}


\begin{thebibliography}{10}

\bibitem{Note1}
{Examples of such applications include the problems of navigation, or
  human-robot interaction. We note that the focus on such classical
  environments is by no means a restriction of our scheme. For a quantum
  environment, the actions which the agent can perform could e.g. be quantum
  measurements (as components of his actuators), and percepts the measurement
  outcomes -- such scenarios are certainly not without interest. The results we
  present in this work apply equally to such environments. However, we will not
  explore the applications of learning agents in quantum environments in this
  paper.}

\bibitem{Note2}
Even if we were to allow superpositions of actions, the amount of control the
  agent must have over the degrees of freedom of the environment, in order to
  apply quantum query algorithms, may be prohibitive. This constitutes one of
  the fundamental operative distinctions between \protect \emph {quantum
  algorithms,} where full control is assumed, and \protect \emph {quantum
  agents,} where such control is limited.

\bibitem{Note3}
In embodied agents, these physical processes may \protect \emph {e.g.} realize
  some internal representation of the environment, which the agent itself has
  to develop as it interacts with the environment. For example, in the context
  of artificial neural networks such internal models are known as
  self-organizing maps and, more specifically, sensorimotor maps \cite
  {1995_Kohonen,2006_Toussaint}.

\bibitem{Note4}
In the PS framework one formally distinguishes between real percepts and
  actions $s,a$, and their internal representations denoted $\mu (s),\mu (a)$,
  which comprise clips. For our purposes, by abuse of notation, we will omit
  the mapping $\mu $. For more details see the Appendix, section \ref
  {abst:ThePSModel}.

\bibitem{Note5}
In the PS, the hopping probabilities themselves are encoded in the so-called
  $h-$matrix, which is an un-normalized representation of the transition matrix
  \cite {2012_Briegel}.

\bibitem{Note6}
Each agent is equipped with input and output couplers, which translate, through
  sensors and actuators (see Fig.~\ref {fig:AgentBasic}), real percepts to the
  internal representations of percepts, and internal representations of actions
  to real actions.

\bibitem{Note7}
In this paper we do not consider logarithmically contributing terms in the
  complexity analysis, thus we use the $\protect \mathaccentV
  {tilde}07E{O}$-level analysis of the limiting behavior (instead of the
  standard 'big O' $O$).

\bibitem{Note8}
The optimality claim holds, provided that no additionaly mechanisms except for
  diffusion and checking are available.

\bibitem{Note9}
In the case of reversible MCs, which will be the main focus of this paper,
  $P=P^*$, so $V$ can be constructed from $U$ by conjugating it with the swap
  operator of the two registers. Here, we present the construction for the
  general case of irreducible, aperiodic Markov chains.

\bibitem{Note10}
Here, we note that concrete applications of quantum walks specified by
  non-symmetric Markov chains have, to our knowledge and aside from this work,
  only been studied in \cite {2012_paparo_google,2013_paparo_complex}, by two
  of the authors and other collaborators, in a significantly different context.

\bibitem{Note11}
For completeness we note as a technicality, following \cite
  {2011_Magniez_SIAM}, that if $\epsilon _s$ is always bounded below by a known
  constant (say $3/4$ as in \cite {1998_Boyer}), the quantum agent can
  immediately measure the initial state and efficiently produce the desired
  output by iterating this process a constant number of times. However, in the
  scenarios we envision, $\epsilon _s$ is very small.

\bibitem{Note12}
We note that some of the only logarithmically contributing terms, which appear
  in a more detailed analysis omitted here, can be further avoided using more
  complicated constructions as has, for instance, been done in \cite
  {2011_Magniez_SIAM}.

\bibitem{Note13}
We acknowledge that the \protect \emph {interpretation} of these results in the
  sense of their practical impact on the field are not without controversy.
  Nonetheless, the validity of the mathematical statements is not contested.
  See \cite {NFLorg} for more details.

\bibitem{Note14}
That is, a comparison can be made with no further assumptions on the
  environment beyond the trivial onesÑthat the percept and action sets are
  compatible with the environment and that the environment provides a rewarding
  scheme.

\bibitem{Note15}
Additionally, the quality of the approximation can be, probabilistically,
  arbitrarily raised by one round of phase estimation algorithm for the walk
  operator, provided no phase is detected. Should this fail, the `unsearching'
  process is repeated.

\end{thebibliography}


\begin{thebibliography}{10}
\expandafter\ifx\csname url\endcsname\relax
  \def\url#1{\texttt{#1}}\fi
\expandafter\ifx\csname urlprefix\endcsname\relax\def\urlprefix{URL }\fi
\providecommand{\bibinfo}[2]{#2}
\providecommand{\eprint}[2][]{\url{#2}}

\item[]{{\normalsize \textbf{References:\\}}}
\bibitem{1985_Deutsch}
\bibinfo{author}{Deutsch, D.}
\newblock \bibinfo{title}{Quantum theory, the {C}hurch-{T}uring principle and
  the universal quantum computer}.
\newblock \emph{\bibinfo{journal}{Proceedings of the Royal Society of London
  A}} \textbf{\bibinfo{volume}{400}}, \bibinfo{pages}{97--117}
  (\bibinfo{year}{1985}).
\newblock \urlprefix\url{http://www.qubit.org/oldsite/resource/deutsch85.pdf}.


\bibitem{1992_Deutsch}
\bibinfo{author}{Deutsch, D.} \& \bibinfo{author}{Jozsa, R.}
\newblock \bibinfo{title}{Rapid solution of problems by quantum computation}.
\newblock \emph{\bibinfo{journal}{Proc Roy Soc Lond A}}
  \textbf{\bibinfo{volume}{439}}, \bibinfo{pages}{553--558}
  (\bibinfo{year}{1992}).

\bibitem{1996_Grover}
\bibinfo{author}{Grover, L.~K.}
\newblock \emph{\bibinfo{journal}{Proceedings, 28th Annual ACM Symposium on the
  Theory of Computing (STOC)}} \bibinfo{pages}{212} (\bibinfo{year}{1996}).

\bibitem{1994_Shor}
\bibinfo{author}{Shor, P.~W.}
\newblock \bibinfo{title}{Algorithms for quantum computation: discrete
  logarithms and factoring}.
\newblock In \emph{\bibinfo{booktitle}{Proceedings of the 35th symposium on
  foundations of computer science, FOCS~'94}}, \bibinfo{pages}{124--134}
  (\bibinfo{publisher}{IEEE}, \bibinfo{year}{1994}).

\bibitem{2000_NC}
\bibinfo{author}{Nielsen, M.~A.} \& \bibinfo{author}{Chuang, I.~L.}
\newblock \emph{\bibinfo{title}{Quantum Computation and Quantum Information}}
  (\bibinfo{publisher}{Cambridge University Press}, \bibinfo{year}{2000}).

\bibitem{2000_Bennet}
\bibinfo{author}{Charles H.~Bennett, D. P.~D.}
\newblock \bibinfo{title}{Quantum information and computation}.
\newblock \emph{\bibinfo{journal}{Nature}} \bibinfo{pages}{247Ð255}
  (\bibinfo{year}{2000}).
\newblock
  \urlprefix\url{http://www.nature.com/nature/journal/v404/n6775/full/404247a0.html}.

\bibitem{2002_Sasaki}
\bibinfo{author}{Sasaki, M.} \& \bibinfo{author}{Carlini, A.}
\newblock \bibinfo{title}{Quantum learning and universal quantum matching
  machine}.
\newblock \emph{\bibinfo{journal}{Phys. Rev. A}} \textbf{\bibinfo{volume}{66}},
  \bibinfo{pages}{022303} (\bibinfo{year}{2002}).
\newblock \urlprefix\url{http://link.aps.org/doi/10.1103/PhysRevA.66.022303}.

\bibitem{2008_Neven}
\bibinfo{author}{Neven, H.}, \bibinfo{author}{Denchev, V.~S.},
  \bibinfo{author}{Rose, G.} \& \bibinfo{author}{Macready, W.~G.}
\newblock \bibinfo{title}{Training a binary classifier with the quantum
  adiabatic algorithm}  (\bibinfo{year}{2008}).
\newblock \eprint{arXiv/0811.0416}.

\bibitem{2013_Lloyd}
\bibinfo{author}{Lloyd, S.}, \bibinfo{author}{Mohseni, M.} \&
  \bibinfo{author}{Rebentrost, P.}
\newblock \bibinfo{title}{Quantum algorithms for supervised and unsupervised
  machine learning}.
\newblock \emph{\bibinfo{journal}{ArXiv:1307.0411}}  (\bibinfo{year}{2013}).

\bibitem{2009_Brukner}
\bibinfo{author}{Manzano, D.}, \bibinfo{author}{Paw{\l}owski, M.} \&
  \bibinfo{author}{Brukner, {\v{C}}.}
\newblock \bibinfo{title}{The speed of quantum and classical learning for
  performing the k th root of not}.
\newblock \emph{\bibinfo{journal}{New Journal of Physics}}
  \textbf{\bibinfo{volume}{11}}, \bibinfo{pages}{113018}
  (\bibinfo{year}{2009}).
\newblock \urlprefix\url{http://stacks.iop.org/1367-2630/11/i=11/a=113018}.

\bibitem{2013_Lidar}
\bibinfo{author}{Pudenz, K.~L.} \& \bibinfo{author}{Lidar, D.~A.}
\newblock \bibinfo{title}{Quantum adiabatic machine learning}.
\newblock \emph{\bibinfo{journal}{Quantum Information Processing}}
  \textbf{\bibinfo{volume}{12}}, \bibinfo{pages}{2027--2070}
  (\bibinfo{year}{2013}).
\newblock \urlprefix\url{http://dx.doi.org/10.1007/s11128-012-0506-4}.

\bibitem{2013_Aimeur}
\bibinfo{author}{A{\"{i}}meur, E.}, \bibinfo{author}{Brassard, G.} \&
  \bibinfo{author}{Gambs, S.}
\newblock \bibinfo{title}{Quantum speed-up for unsupervised learning}.
\newblock \emph{\bibinfo{journal}{Machine Learning}}
  \textbf{\bibinfo{volume}{90}}, \bibinfo{pages}{261--287}
  (\bibinfo{year}{2013}).
\newblock \urlprefix\url{http://dx.doi.org/10.1007/s10994-012-5316-5}.

\bibitem{1986_Braitenberg}
\bibinfo{author}{Braitenberg, V.}
\newblock \emph{\bibinfo{title}{Vehicles: Experiments in Synthetic Psychology}}
  (\bibinfo{publisher}{MIT Press}, \bibinfo{address}{Cambridge Massachusetts},
  \bibinfo{year}{1986}).

\bibitem{1999_Brooks}
\bibinfo{author}{Brooks, R.}
\newblock \emph{\bibinfo{title}{Cambrian intelligence: The early history of the
  new AI}} (\bibinfo{publisher}{The MIT Press}, \bibinfo{address}{Cambridge,
  MA}, \bibinfo{year}{1999}).

\bibitem{1999_Pfeifer}
\bibinfo{author}{Pfeifer, R.} \& \bibinfo{author}{Scheier, C.}
\newblock \emph{\bibinfo{title}{Understanding Intelligence}}
  (\bibinfo{publisher}{MIT Press}, \bibinfo{address}{Cambridge, MA, USA},
  \bibinfo{year}{1999}).

\bibitem{2006_Pfeifer}
\bibinfo{author}{Pfeifer, R.} \& \bibinfo{author}{Bongard, J.~C.}
\newblock \emph{\bibinfo{title}{How the Body Shapes the Way We Think: A New
  View of Intelligence (Bradford Books)}} (\bibinfo{publisher}{The MIT Press},
  \bibinfo{year}{2006}).

\bibitem{2008_Floreano}
\bibinfo{author}{Floreano, D.} \& \bibinfo{author}{Mattiussi, C.}
\newblock \emph{\bibinfo{title}{Bio-inspired artificial intelligence :
  theories, methods, and technologies}} (\bibinfo{publisher}{The MIT Press},
  \bibinfo{year}{2008}).



\bibitem{2008_Barsalou}
\bibinfo{author}{Barsalou, L.~W.}
\newblock \bibinfo{title}{Grounded cognition}.
\newblock \emph{\bibinfo{journal}{Annual Review of Psychology}}
  \textbf{\bibinfo{volume}{59}}, \bibinfo{pages}{617--645}
  (\bibinfo{year}{2008}).

\bibitem{2012_Briegel}
\bibinfo{author}{Briegel, H.~J.} \& \bibinfo{author}{De~las Cuevas, G.}
\newblock \bibinfo{title}{Projective simulation for artificial intelligence}.
\newblock \emph{\bibinfo{journal}{Sci. Rep.}} \textbf{\bibinfo{volume}{2}}
  (\bibinfo{year}{2012}).

\bibitem{2003_Russel}
\bibinfo{author}{Russel, S.~J.} \& \bibinfo{author}{Norvig, P.}
\newblock \emph{\bibinfo{title}{Artificial intelligence - A modern approach}}
  (\bibinfo{publisher}{Prentice Hall}, \bibinfo{address}{New Jersey},
  \bibinfo{year}{2003}), \bibinfo{edition}{second edition}.




\bibitem{SuttonBarto98}
\bibinfo{author}{Sutton, R.~S.} \& \bibinfo{author}{Barto, A.~G.}
\newblock \emph{\bibinfo{title}{Reinforcement learning: An introduction}}
  (\bibinfo{publisher}{MIT Press, Cambridge Massachusetts},
  \bibinfo{year}{1998}), \bibinfo{edition}{first} edn.

\bibitem{2004_Szegedy_IEEE}
\bibinfo{author}{Szegedy, M.}
\newblock \bibinfo{title}{Quantum speed-up of markov chain based algorithms}.
\newblock In \emph{\bibinfo{booktitle}{Foundations of Computer Science, 2004.
  Proceedings. 45th Annual IEEE Symposium on}}, \bibinfo{pages}{32 -- 41}
  (\bibinfo{year}{2004}).

\bibitem{2011_Magniez_SIAM}
\bibinfo{author}{Magniez, F.}, \bibinfo{author}{Nayak, A.},
  \bibinfo{author}{Roland, J.} \& \bibinfo{author}{Santha, M.}
\newblock \bibinfo{title}{Search via quantum walk}.
\newblock \emph{\bibinfo{journal}{SIAM Journal on Computing}}
  \textbf{\bibinfo{volume}{40}}, \bibinfo{pages}{142--164}
  (\bibinfo{year}{2011}).
\newblock \urlprefix\url{http://epubs.siam.org/doi/abs/10.1137/090745854}.


\bibitem{1995_Kohonen}
\bibinfo{author}{Kohonen, T.}
\newblock \bibinfo{title}{Self-organizing maps}
  (\bibinfo{publisher}{Springer-Verlag}, \bibinfo{year}{1995}).

\bibitem{2006_Toussaint}
\bibinfo{author}{Toussaint, M.}
\newblock \bibinfo{title}{A sensorimotor map: Modulating lateral interactions
  for anticipation and planning}.
\newblock \emph{\bibinfo{journal}{Neural Comput.}}
  \textbf{\bibinfo{volume}{18}}, \bibinfo{pages}{1132--1155}
  (\bibinfo{year}{2006}).
\newblock \urlprefix\url{http://dx.doi.org/10.1162/089976606776240995}.

\bibitem{2005_Magniez}
\bibinfo{author}{Magniez, F.}, \bibinfo{author}{Santha, M.} \&
  \bibinfo{author}{Szegedy, M.}
\newblock \bibinfo{title}{Quantum algorithms for the triangle problem}.
\newblock In \emph{\bibinfo{booktitle}{Proceedings of the Sixteenth Annual
  ACM-SIAM Symposium on Discrete Algorithms}}, SODA '05,
  \bibinfo{pages}{1109--1117} (\bibinfo{publisher}{Society for Industrial and
  Applied Mathematics}, \bibinfo{address}{Philadelphia, PA, USA},
  \bibinfo{year}{2005}).
\newblock \urlprefix\url{http://dl.acm.org/citation.cfm?id=1070432.1070591}.

\bibitem{2004_Ambainis}
\bibinfo{author}{Ambainis, A.}
\newblock \bibinfo{title}{Quantum walk algorithms for element distinctness}.
\newblock In \emph{\bibinfo{booktitle}{45th Annual IEEE Symposium on
  Foundations of Computer Science, OCT 17-19, 2004. IEEE Computer Society
  Press, Los Alamitos, CA}}, \bibinfo{pages}{22--31} (\bibinfo{year}{2004}).

\bibitem{2012_Magniez_Algorithmica}
\bibinfo{author}{Magniez, F.}, \bibinfo{author}{Nayak, A.},
  \bibinfo{author}{Richter, P.~C.} \& \bibinfo{author}{Santha, M.}
\newblock \bibinfo{title}{On the hitting times of quantum versus random walks}.
\newblock \emph{\bibinfo{journal}{Algorithmica}} \textbf{\bibinfo{volume}{63}},
  \bibinfo{pages}{91--116} (\bibinfo{year}{2012}).
\newblock \urlprefix\url{http://dx.doi.org/10.1007/s00453-011-9521-6}.

\bibitem{2011_Temme_Nature}
\bibinfo{author}{Temme, K.}, \bibinfo{author}{Osborne, T.~J.},
  \bibinfo{author}{Vollbrecht, K. G.~H.}, \bibinfo{author}{Poulin, D.} \&
  \bibinfo{author}{Verstraete, F.}
\newblock \bibinfo{title}{Quantum metropolis sampling}.
\newblock\bibinfo{journal}{Nature}
  \textbf{\bibinfo{volume}{471}}, \bibinfo{pages}{87--90}
  (\bibinfo{year}{2011}).

\bibitem{2012_Yung}
\bibinfo{author}{Yung, M.-H.} \& \bibinfo{author}{Aspuru-Guzik, A.}
\newblock \bibinfo{title}{A quantum-quantum Metropolis algorithm}.
\newblock \emph{\bibinfo{journal}{Proc. Natl. Acad. Sci.}}  (\bibinfo{year}{2012}).
\newblock\bibinfo{doi}{10.1073/pnas.1111758109
}

\bibitem{2008_Somma_PRL}
\bibinfo{author}{Somma, R.~D.}, \bibinfo{author}{Boixo, S.},
  \bibinfo{author}{Barnum, H.} \& \bibinfo{author}{Knill, E.}
\newblock \bibinfo{title}{Quantum simulations of classical annealing
  processes}.
\newblock \emph{\bibinfo{journal}{Phys. Rev. Lett.}}
  \textbf{\bibinfo{volume}{101}}, \bibinfo{pages}{130504}
  (\bibinfo{year}{2008}).
\newblock
  \urlprefix\url{http://link.aps.org/doi/10.1103/PhysRevLett.101.130504}.

\bibitem{2008_Wocjan}
\bibinfo{author}{Wocjan, P.} \& \bibinfo{author}{Abeyesinghe, A.}
\newblock \bibinfo{title}{Speedup via quantum sampling}.
\newblock \emph{\bibinfo{journal}{Phys. Rev. A}} \textbf{\bibinfo{volume}{78}},
  \bibinfo{pages}{042336} (\bibinfo{year}{2008}).
\newblock \urlprefix\url{http://link.aps.org/doi/10.1103/PhysRevA.78.042336}.

\bibitem{2012_paparo_google}
\bibinfo{author}{Paparo, G.~D.} \& \bibinfo{author}{Martin-Delgado, M.}
\newblock \bibinfo{title}{Google in a quantum network}.
\newblock \emph{\bibinfo{journal}{Scientific reports}}
  \textbf{\bibinfo{volume}{2}} (\bibinfo{year}{2012}).

\bibitem{2013_paparo_complex}
\bibinfo{author}{Paparo, G.}, \bibinfo{author}{Mueller, M.},
  \bibinfo{author}{Comellas, F.} \& \bibinfo{author}{Martin-Delgado, M.}
\newblock \bibinfo{title}{Quantum google in a complex network}.
\newblock \emph{\bibinfo{journal}{Scientific reports}}
  \textbf{\bibinfo{volume}{3}} (\bibinfo{year}{2013}).

\bibitem{1998_Boyer}
\bibinfo{author}{Boyer, M.}, \bibinfo{author}{Brassard, G.},
  \bibinfo{author}{Hoeyer, P.} \& \bibinfo{author}{Tapp, A.}
\newblock \bibinfo{title}{Tight bounds on quantum searching}.
\newblock \emph{\bibinfo{journal}{Fortsch. Phys.}}
  \textbf{\bibinfo{volume}{46}}, \bibinfo{pages}{493--506}
  (\bibinfo{year}{1996}).
\newblock \eprint{arXiv:quant-ph/9605034}.

\bibitem{2012_Aspuru}
\bibinfo{author}{Aspuru-Guzik, A.} \& \bibinfo{author}{Walther, P.}
\newblock \bibinfo{title}{{Photonic quantum simulators}}.
\newblock \emph{\bibinfo{journal}{Nature Physics}}
  \textbf{\bibinfo{volume}{8}}, \bibinfo{pages}{285--291}
  (\bibinfo{year}{2012}).
\newblock \urlprefix\url{http://dx.doi.org/10.1038/nphys2253}.

\bibitem{2012_Roos}
\bibinfo{author}{Blatt, R.} \& \bibinfo{author}{Roos, C.~F.}
\newblock \bibinfo{title}{{Quantum simulations with trapped ions}}.
\newblock \emph{\bibinfo{journal}{Nature Physics}}
  \textbf{\bibinfo{volume}{8}}, \bibinfo{pages}{277--283}
  (\bibinfo{year}{2012}).
\newblock \urlprefix\url{http://dx.doi.org/10.1038/nphys2252}.

\bibitem{2013_Schaetz}
\bibinfo{author}{Schaetz, T.}, \bibinfo{author}{Monroe, C.~R.} \&
  \bibinfo{author}{Esslinger, T.}
\newblock \bibinfo{title}{Focus on quantum simulation}.
\newblock \emph{\bibinfo{journal}{New Journal of Physics}}
  \textbf{\bibinfo{volume}{15}}, \bibinfo{pages}{085009}
  (\bibinfo{year}{2013}).
\newblock \urlprefix\url{http://stacks.iop.org/1367-2630/15/i=8/a=085009}.

\bibitem{1996_Wolpert}
\bibinfo{author}{Wolpert, D.~H.}
\newblock \bibinfo{title}{The lack of a priori distinctions between learning
  algorithms}.
\newblock \emph{\bibinfo{journal}{Neural Comput.}}
  \textbf{\bibinfo{volume}{8}}, \bibinfo{pages}{1341--1390}
  (\bibinfo{year}{1996}).
\newblock \urlprefix\url{http://dx.doi.org/10.1162/neco.1996.8.7.1341}.

\bibitem{1997_Droste}
\bibinfo{author}{Droste, S.}, \bibinfo{author}{Jansen, T.} \&
  \bibinfo{author}{Wegener, I.}
\newblock \bibinfo{title}{Optimization with randomized search heuristics -- the
  (a)nfl theorem, realistic scenarios, and difficult functions}.
\newblock \emph{\bibinfo{journal}{Theoretical Computer Science}}
  \textbf{\bibinfo{volume}{287}}, \bibinfo{pages}{2002} (\bibinfo{year}{1997}).

\bibitem{NFLorg}
\bibinfo{title}{Introduction to discussions regarding the no free lunch
  theorems, and their interpretations, can be found at the following web page}.
\newblock \bibinfo{howpublished}{\url{http://no-free-lunch.org/}}.

\bibitem{1993_Sinclair}
\bibinfo{author}{Sinclair, A.}
\newblock \emph{\bibinfo{title}{Algorithms for random generation and counting:
  a Markov chain approach}}, vol.~\bibinfo{volume}{7}
  (\bibinfo{publisher}{Springer}, \bibinfo{year}{1993}).

\bibitem{Julian12}
\bibinfo{author}{Mautner, J.}, \bibinfo{author}{Makmal, A.},
  \bibinfo{author}{Manzano, D.}, \bibinfo{author}{Tiersch, M.} \&
  \bibinfo{author}{Briegel, H.~J.}
\newblock \bibinfo{title}{Projective simulation for classical learning agents:
  a comprehensive investigation.} \bibinfo{note}{To appear in New Generation Comput.}.

\bibitem{Briegel2_2012}
\bibinfo{author}{Briegel, H.}
\newblock \bibinfo{title}{{On creative machines and the physical origins of
  freedom.}}
\newblock \emph{\bibinfo{journal}{Scientific reports}}
  \textbf{\bibinfo{volume}{2}}, \bibinfo{pages}{522} (\bibinfo{year}{2012}).
\newblock \urlprefix\url{http://dx.doi.org/10.1038/srep00522}.

\bibitem{2007_Magniez_SIAM}
\bibinfo{author}{Magniez, F.}, \bibinfo{author}{Santha, M.} \&
  \bibinfo{author}{Szegedy, M.}
\newblock \bibinfo{title}{Quantum algorithms for the triangle problem}.
\newblock \emph{\bibinfo{journal}{SIAM Journal on Computing}}
  \textbf{\bibinfo{volume}{37}}, \bibinfo{pages}{413--424}
  (\bibinfo{year}{2007}).
\newblock \urlprefix\url{http://epubs.siam.org/doi/abs/10.1137/050643684}.




\bibitem{1993_Vazirani}
\bibinfo{author}{Bernstein, E.} \& \bibinfo{author}{Vazirani, U.}
\newblock \bibinfo{title}{Quantum complexity theory}.
\newblock In \emph{\bibinfo{booktitle}{in Proc. 25th Annual ACM Symposium on
  Theory of Computing, ACM}}, \bibinfo{pages}{11--20} (\bibinfo{year}{1993}).


\end{thebibliography}
\end{document}